\documentclass[12pt]{article}
\usepackage{color}
\usepackage{graphicx}

\usepackage{amsmath, amsthm, amssymb}
\newtheorem{alg}{Algorithm}

\usepackage{accents}
\newcommand*{\dt}[1]{%
  \accentset{\mbox{\large\bfseries .}}{#1}}

\newcommand{\RR}{\mathbb R}

\title{Data assimilation in the low noise regime\\  with application to the Kuroshio}
\author{Eric Vanden-Eijnden\footnotemark[1]~  and Jonathan Weare\footnotemark[2]}
\begin{document}

\footnotetext[1]{The Courant Institute, New York University}
\footnotetext[2]{Department of Mathematics, University of Chicago}

\maketitle

\begin{abstract}
On-line data assimilation techniques such as ensemble Kalman filters and particle filters lose accuracy dramatically when presented with an unlikely observation.  Such an observation may be caused by an unusually large measurement error or reflect a rare fluctuation in the dynamics of the system.  Over a long enough span of time it becomes likely that one or several of these events will occur.  Often they are signatures of the most interesting features of the underlying system and their prediction becomes the primary focus of the data assimilation procedure.  The Kuroshio or Black Current that runs along the eastern coast of Japan is an example of such a system.  It undergoes infrequent but dramatic changes of state between a small meander during which the current remains close to the coast of Japan, and a large meander during which it bulges away from the coast. Because of the important role that the Kuroshio plays in distributing heat and salinity in the surrounding region, prediction of these transitions is of acute interest.  Here we focus on a regime in which both the
stochastic forcing on the system and the observational noise are small.  In this setting large deviation theory can be used to understand why standard filtering methods fail and guide the design of the more effective data assimilation techniques.   Motivated by our analysis we propose several data assimilation strategies capable of efficiently handling rare events such as the transitions of the Kuroshio.  These techniques are tested on a model of the Kuroshio and shown to perform much better than standard filtering methods.
\end{abstract}

%
%%%%%%%%%%%%%%%%%%%%%%%%%%%%%%%%%%%%%%%%%%%%%%%%%%%%%%%%%%%%%%%%%%%%%
% TITLE
%
% Enter your TITLE here
%%%%%%%%%%%%%%%%%%%%%%%%%%%%%%%%%%%%%%%%%%%%%%%%%%%%%%%%%%%%%%%%%%%%%

%
% Author names, with corresponding author information. 
% [Update and move the \thanks{...} block as appropriate.]
%

%
%\ifthenelse{\boolean{dc}}
%{
%\twocolumn[
%\begin{@twocolumnfalse}
%\amstitle

%
%\begin{center}
%\begin{minipage}{13.0cm}
%\begin{abstract}
%	\myabstract
%	\newline
%	\begin{center}
%		\rule{38mm}{0.2mm}
%	\end{center}
%\end{abstract}
%\end{minipage}
%\end{center}
%\end{@twocolumnfalse}
%]
%}
%{
%\amstitle
%\begin{abstract}
%\myabstract
%\end{abstract}
%\newpage
%}
%%%%%%%%%%%%%%%%%%%%%%%%%%%%%%%%%%%%%%%%%%%%%%%%%%%%%%%%%%%%%%%%%%%%%
% MAIN BODY OF PAPER
%%%%%%%%%%%%%%%%%%%%%%%%%%%%%%%%%%%%%%%%%%%%%%%%%%%%%%%%%%%%%%%%%%%%%
%
\section{Introduction}
\label{intro}
The assimilation of noisy observations into a model to improve its
predictive capabilities is a recurring challenge in many applications.
Examples include weather prediction and forecasting, robot tracking,
stochastic volatility estimation, image analysis, etc. (see
\cite{defreitas05}). In these applications and many others one is
interested in predicting how the system evolves in time given a model
for its dynamics and sequentially available, incomplete observations of its
state.  For practical reasons it is desirable to
assimilate the observations
in real time (``on-line'') via a recursive algorithm that requires
only the latest observation together with the previous estimate of the
system's state. In the simplest case
of Gaussian (i.e. linear) evolution and Gaussian observations, a
solution to this problem is given by the Kalman filter (see
\cite{kalman60}) and extensions thereof (see
\cite{evensen03,ott2004_kalman_localization,hunt2007}), which predict
how the mean and the variance of the system's state evolve given the
observations.  For problems with significantly non-Gaussian features,
Kalman filters are unsuitable (see \cite{miller:1994:jas,miller:1999:tellus,bickel2010_enkal,
  law12} and the numerical results in Section
\ref{sec:model}.\ref{sec:results}), and particle filters, first
suggested in \cite{gordon93} and \cite{kitagawa96}, can be used
instead. Particle filters, also known as sequential Monte Carlo
methods, are recursive assimilation algorithms that predict how the
posterior distribution of the system's state evolves given the
observations. They can in principle be applied to general, nonlinear,
non-Gaussian situations, though they can be impractical in high
dimensional problems (see \cite{bickel2009_nleaf}).  An additional problem
that both Kalman filters and particle filters share is that they tend
to fail when the system undergoes occasional, unusually large
transitions revealed by an observation that is inconsistent with the
predictive distribution of the system's state. Over long periods of
time such transitions are inevitable and in some systems they are
precisely the events of main interest. Our aim here is to { provide 
a clear mathematical description of the failure of standard filtering techniques and, 
guided by that analysis, propose  methods that remain accurate in the presense of 
rare events.}

\begin{figure}[t]\noindent
\includegraphics[width=33pc,angle=0]{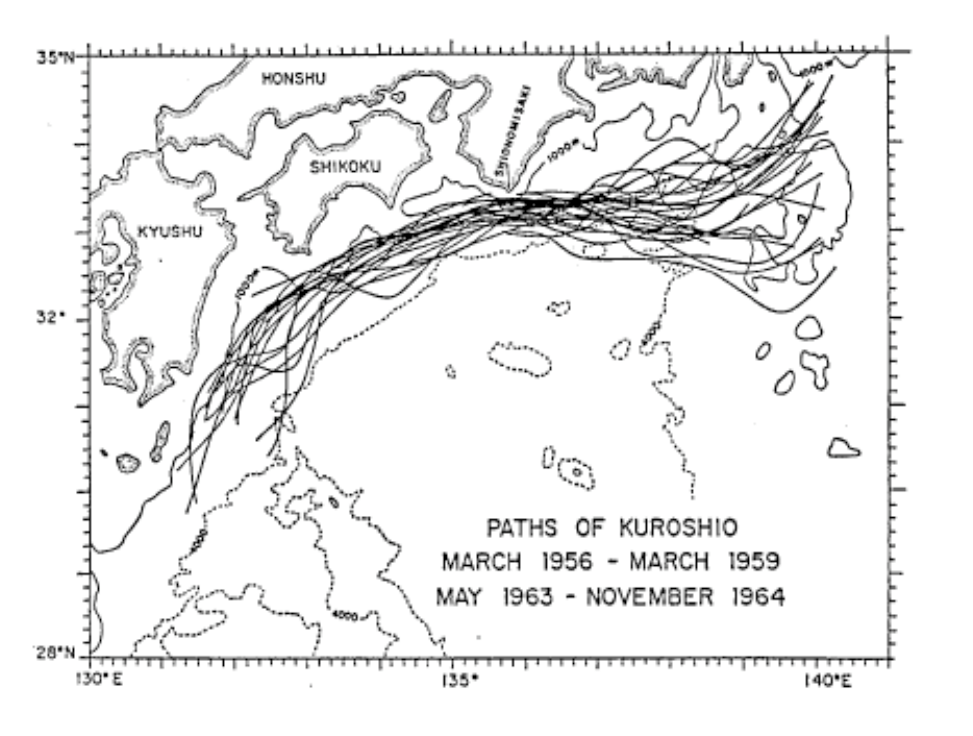}
\caption{ Paths in the small meander state.  
 (Reproduced from \cite{qiu00}.  Originally adapted from \cite{taft72}.)}
\label{smallmeander}
\end{figure}

\begin{figure}[t]
\noindent
\includegraphics[width=33pc,angle=0]{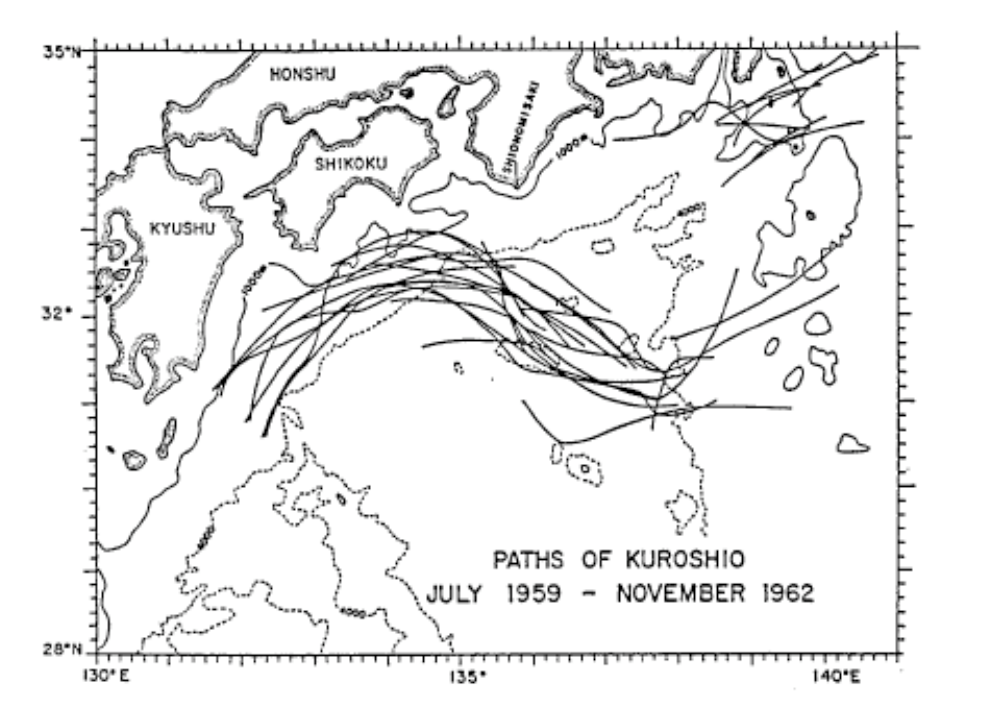}
\caption{Paths in the large meander state.  (Reproduced from \cite{qiu00}.  Originally adapted from \cite{taft72}.)}
\label{largemeander}
\end{figure}

An interesting example of a system exhibiting very occasional but
interesting transitions that present significant challenges to
standard data assimilation strategies is the Kuroshio running
along the eastern coast of Japan.  The Kuroshio exhibits
transitions between a small meander during which the current remains
close to the coast of Japan, and a large meander during which the
current bulges away from the coast (see Figures \ref{smallmeander} and
\ref{largemeander}).  The Kuroshio's central role in distributing heat
and salinity in the surrounding region has led to many studies of its
bimodal behavior beginning with a study by Yoshida in 1961 (see
\cite{yoshida61}).  The meanders typically persist for 5-10 years,
while the transitions between them occur in only a few months. Here we
focus on a simple model qualitatively capturing the bimodal behavior
of the Kuroshio as a test bed for our new filtering strategies, and we
show that they remain effective in a regime where standard filtering
techniques fail dramatically. For data assimilation studies using more accurate 
models of the Kuroshio see, for example, \cite{Miyazawa:2008:jgr} and the references therein. 
 Our general statements and results apply
to systems exhibiting less dramatic and less interesting rare events,
as these also occur eventually and lead to a loss of accuracy in
standard filtering techniques.

We shall focus primarily on situations where the system is forced by a
small stochastic term and the observational noise is small, as this
regime is the most challenging for standard filters. { The stochastic forcing in the dynamics can be a consequence of small scale physical forcings or an attempt to represent model errors. Even when our faith in the model is low and one adds  a significant stochastic forcing to represent that uncertainty  one must keep in mind that in  any reasonable model of a system like the Kuroshio the rare event of interest will remain infrequent.   Our low noise}, accurate observation regime, allows the behavior of rare events such
as the transitions of the Kuroshio to be understood within the
framework of large deviations theory (see
\cite{Dembo:1993p6678,dupuisellisbook,Freidlin:1984p4427,Varadhan:1985p6658}).
This theory is built on the key observation that, when a rare event
occurs, it typically does so in a predicable way by the most likely
path possible. While the event is rare and the likelihood of observing this path
is
small, the likelihood of observing any other
path is much smaller. The identification of the most likely path
is at the core of a family of data assimilation techniques, called
variational methods (e.g. 3DVar and 4DVar), that do effectively handle
non-linearities in the underlying system (see e.g. \cite{Lewis:2006:book}).
These methods do not involve any sampling.  Instead they find the
solution to a large optimization problem which gives either the most
likely current state of the system given the observations or the most
likely path of the system over several observational windows.
Without modification these methods provide only an estimate of the most
likely trajectory of the system given the observations and do not
provide further information about the distribution of the system.   Modifications
can yeild some limited (and sometimes very unreliable) information about the 
local variation around the most likely trajectory.   Moreover,
they are not directly suitable for on-line data assimilation.

Several authors have suggested combining sampling methods with
variational methods (see e.g. \cite{yang2009} and references
therein). The framework we suggest in this paper follows this line of
thought and naturally leads to hybrid data assimilation techniques
sharing characteristics of both sampling based filters and variational
methods. More specifically, we use information about the most likely
path for the event to do importance sampling within the particle
filters. This approach not only leads to filters that remain accurate
in the presence of rare events, but it also allows one to predict the
behavior of new and existing data assimilation schemes.  We
demonstrate that, in the small noise regime that is our focus,
standard schemes can be expected to behave very poorly when the
underlying system undergoes a rare large change in its state.  The
framework and methods suggested here should also shed some light on
interesting and related methods such as path sampling based techniques (see
\cite{alexander2004,apte08, weare09}), ``nudging'' techniques (see \cite{Stauffer:1993:TellusA}),   implicit sampling
(see \cite{chorin2009_filter1,chorin2009_filter2}), and sequential importance sampling schemes (see \cite{vanLeeuwen:2010:RMetS}).

The remainder of this paper is organized as follows. In
Section~\ref{sec:nlfilter} we describe the filtering problem along
with the most basic versions of the Kalman filter
(Section~\ref{sec:nlfilter}.\ref{sec:kalman}) and a particle filter
(Section~\ref{sec:nlfilter}.\ref{sec:particle}).  In
Section~\ref{sec:small noise} we discuss the difficulties that these
filters encounter in the small noise, accurate observation regime. In
Section~\ref{sec:importance}, we propose several hybrid sampling
strategies. Section~\ref{sec:model} presents the results of our
numerical experiments on a simple model of the Kuroshio. Some
conclusions are given in Section~\ref{sec:conclusion}. Finally, in
Appendices A and B we give details of some aspects of the numerical
implementation of our algorithms.

\section{Ensemble filtering methods}
\label{sec:nlfilter}

The goal of any discrete-time filtering algorithm is to approximately
reconstruct the trajectory of some time-dependent process $\mathbf{x}(t)$ from
observations $\mathbf{y}^\circ(t_1)$, $\mathbf{y}^\circ(t_2)$, ... taken at a discrete set of times
$t_1$, $t_2,$ ... The observations are typically incomplete and made
with measurement error, i.e. they are of the form
\[
\mathbf{y}^\circ(t_n) = H\left( \mathbf{x}(t_n)\right) + \xi_n
\]
where the $H(x)$ is some function of the state space and $\xi_n$ models observation error (note that the dimension of $\mathbf{y}^\circ$ may be much lower than the dimension of $\mathbf{x}$).  We
will assume that, conditioned on $\mathbf{x}(t_n)$ the
observations, $\mathbf{y}^\circ(t_n),$ admit a probability density function
\[
p(\mathbf{y}^\circ(t_n)| \mathbf{x}(t_n)).
\] 
When $\xi$ is gaussian we have
\[
p(\mathbf{y}^\circ(t_n)| \mathbf{x}(t_n)) \propto \exp\left(-\frac{1}{2} (\mathbf{y}^\circ(t_n)- H(\mathbf{x}(t_n)))^\text{\tiny T} \mathbf{R}^{-1} (\mathbf{y}^\circ(t_n)- H(\mathbf{x}(t_n)))\right).
\]
where $\mathbf{R}$ is the covariance matrix of $\xi_n.$
% here and below we follow standard practice and label random
%variables with upper case letters (e.g. $\mathbf{x}(t_n)$, $H_n$, etc.) and
%the values assumed by those random numbers with lower case letters (e.g. $\mathbf{x}(t_n)$,
%$\mathbf{y}^\circ(t_n)$, etc.).
The underlying process $\mathbf{x}(t)$ should be considered ``hidden''
and revealed only through the observations. Ideally one would like to
 calculate modes and moments of the conditional distribution
of the hidden signal $\mathbf{x}(t)$ given these observations.  For example, one
may wish to approximate the expectation of $f(\mathbf{x}(t_n))$ conditional on
$\mathbf{y}^\circ(t_1),\dots, \mathbf{y}^\circ(t_n)$,
\begin{equation}\label{condexp}
\mathbf{E}\left[ f(\mathbf{x}(t_n))\, \big\vert\, \mathbf{y}^\circ(t_1),\dots,\mathbf{y}^\circ(t_n) \right]
\end{equation}

When the underlying process $\mathbf{x}(t)$ is Markovian there are several
recursive approaches to the approximation of quantities such as
\eqref{condexp}. These methods are based on the relationship
\begin{multline}\label{filterrecur}
  p\left(\mathbf{x}(t_n)\, \big\vert\, \mathbf{y}^\circ(t_1),\dots,\mathbf{y}^\circ(t_n) \right) =\\ \frac{p\left(\mathbf{y}^\circ(t_n) | \mathbf{x}(t_n)
  \right) \int p(\mathbf{x}(t_n)\, \vert\, \mathbf{x}(t_{n-1}) ) p\left(\mathbf{x}(t_{n-1}) \,\big\vert\,
    \mathbf{y}^\circ(t_1),\dots,\mathbf{y}^\circ(t_{n-1}) \right)dx_{n-1}}{p(\mathbf{y}^\circ(t_n)\,|\,\mathbf{y}^\circ(t_1),\dots,\mathbf{y}^\circ(t_{n-1}))}
\end{multline}
which follows from the Markov property of $\mathbf{x}(t)$ and Bayes' Formula.
Here $p(\mathbf{x}(t_n)\,|\, \mathbf{y}^\circ(t_1),\dots,\mathbf{y}^\circ(t_n) ),$ denotes the probability density
function of $\mathbf{x}(t_n)$ conditional on $\mathbf{y}^\circ(t_1),\dots, \mathbf{y}^\circ(t_n)$ and
will be referred to as the posterior density, and $p(\mathbf{x}(t_n)\, \vert\,
\mathbf{x}(t_{n-1}) ),$ denotes the density of $\mathbf{x}(t_n)$ conditional on
$\mathbf{x}(t_{n-1})$ and will be referred to as the predictive density.
The interpretation of equation~\eqref{filterrecur} is simple: it
states that to obtain the posterior density at time $t_n$ from the
posterior density at time $t_{n-1}$ one can first integrate the system
from time $t_{n-1}$ to time $t_n$ with initial conditions drawn from
the posterior density at time $t_{n-1}$ and then discount the
resulting samples by a weight proportional to $p\left( \mathbf{y}^\circ(t_n) |
  \mathbf{x}(t_n)\right).$
  In fact, 
if $\mathbf{x}(t_{n-1})$ is assumed to be drawn from 
$p\left(\mathbf{x}(t_{n-1})\, \big\vert\,  \mathbf{y}^\circ(t_1),\dots,\mathbf{y}^\circ(t_{n-1}) \right),$ then
\[
\mathbf{E}\left[ f(\mathbf{x}(t_n))\, \big\vert\, \mathbf{y}^\circ(t_1),\dots,\mathbf{y}^\circ(t_n) \right] = \frac{\mathbf{E}\left[f(\mathbf{x}(t_n))p(\mathbf{y}^\circ(t_n)|\mathbf{x}(t_n))\right]}{\mathbf{E}\left[p(\mathbf{y}^\circ(t_n) |\mathbf{x}(t_n))\right]}
\]

There are several methods by which one might hope to
approximately carry out the recursion \eqref{filterrecur}.  Perhaps
the most obvious approach is to simply compute the integral using some
quadrature scheme.  This approach suffers from two insurmountable
difficulties.  The first is that there is often no closed form
expression for the predictive density $p(\mathbf{x}(t_n)\,\vert\, \mathbf{x}(t_{n-1})).$ The
second is that numerical quadrature becomes computationally
impractical in more than a few dimensions.  Two popular approximation
methods that have been applied successfully in various settings are
Kalman and particle filters.  The next two subsections briefly
describe these two approaches.

\subsection{Kalman filter}
\label{sec:kalman}

When the $\xi_n$ are Gaussian random variables and  both the initial density and the predictive density
${p}(\mathbf{x}(t_n)\, \vert\, \mathbf{x}(t_{n-1}) )$ are Gaussian (which is true if the
evolution of $\mathbf{x}(t)$ is governed by a linear equation), it is easy to
see from \eqref{filterrecur} that the posterior density $
{p}\left(\mathbf{x}(t_n)\, \big\vert\, \mathbf{y}^\circ(t_1),\dots,\mathbf{y}^\circ(t_n) \right)$ is also Gaussian.
In this case to completely describe the posterior density one needs
only find formulas for its mean and variance, which are obtained by
easy manipulations of $p(\mathbf{y}^\circ(t_n)\,|\,\mathbf{x}(t_n)),$ ${p}(\mathbf{x}(t_n)\, \vert\, \mathbf{x}(t_{n-1}) ),$
and the initial density.  In fact, from \eqref{filterrecur} one can
derive recursive formulas for the mean and the variance of $
{p}\left(\mathbf{x}(t_n)\, \big\vert\, \mathbf{y}^\circ(t_1),\dots,\mathbf{y}^\circ(t_n) \right).$ The resulting
recursive scheme is called the Kalman filter (see \cite{kalman60}).

There are several important derivatives of the Kalman filter.  These
methods use Gaussian approximations of the densities appearing in
\eqref{filterrecur} ($p(\mathbf{y}^\circ(t_n)\,|\,\mathbf{x}(t_n)),$ 
${p}(\mathbf{x}(t_n)\,
\vert\, \mathbf{x}(t_{n-1}) ),$ and the initial condition) and are accurate in regimes in
which these
densities are nearly Gaussian.  A particularly effective method of this type is the
ensemble Kalman filter (see \cite{evensen03}).  In Algorithm \ref{ek1} below we describe a very simple version of an ensemble Kalman filter.   We will refer to it as an ensemble Kalman filter though production level ensemble Kalman filters may bear it little resemblance.
 We proceed from an
estimate of the mean $\mathbf{x}^a_{n-1}$ and covariance $\widehat\Sigma_{n-1}$
of $ {p}\left(\mathbf{x}(t_{n-1})\, \big\vert\, \mathbf{y}^\circ(t_1),\dots,\mathbf{y}^\circ(t_{n-1}) \right)$ as
follows
\begin{alg}[Basic ensemble Kalman filter]
\label{ek1}
\item[]
{\tt \begin{enumerate}
  \item Generate $M$ independent Gaussian vectors,
    $\left\{\mathbf{x}_j(t_{n-1})\right\}_{j=1}^M,$ with mean \linebreak  $\mathbf{x}^a_{n-1}$
    and covariance $\mathbf{P}^a_{n-1}.$
  \item For each $j,$ evolve $\mathbf{x}_j(t_{n-1})$ from time $t_{n-1}$ to
    time $t_n$ to generate an \linebreak  independent sample $\widetilde {\mathbf{x}}_j(t_n)$
    from ${p}(\mathbf{x}(t_n)\, \vert\, \mathbf{x}_j(t_{n-1}) ).$ Note that, in general,\linebreak
    these samples are no longer Gaussian.
  \item Compute the sample mean $\mathbf{x}^f_n$ and sample covariance $\mathbf{P}^f_n$
    of $\left\{\widetilde{\mathbf{x}}_j(t_n)\right\}_{j=1}^M.$
  \item Analytically compute the mean $\mathbf{x}^a_n$ and covariance $
    \mathbf{P}^a_n$ of the posterior\linebreak Gaussian density 
\[
 C^{-1} p(\mathbf{y}^\circ(t_n)\,|\,\mathbf{x}(t_n)) e^{-\frac12(\mathbf{x}(t_n)-\mathbf{x}^f_n)^\text{\tiny T} \mathbf{P}_n^{-1}(\mathbf{x}(t_n)-\mathbf{x}^f_n)}.
\]
where $C$ is a normalization factor.
\end{enumerate}
}
\end{alg}

\subsection{Particle filter}
\label{sec:particle}

Particle filtering is a very general sequential importance sampling
implementation of the recursion in \eqref{filterrecur} (see
\cite{gordon93,kitagawa96,miller99,liu02,defreitas05}).  Assuming that
one has samples approximately generated from the posterior density at
time $t_{n-1},$ one first evolves the samples forward to time $t_n$
(as in the ensemble Kalman filter).  At this point one has samples
approximately drawn from
\[
{p}\left(\mathbf{x}(t_n)\,\vert \mathbf{y}^\circ(t_1),\dots,\mathbf{y}^\circ(t_{n-1})\right) =  \int{p}(\mathbf{x}(t_n)\, \vert\, \mathbf{x}(t_{n-1}) ) 
 {p}\left(\mathbf{x}(t_{n-1})\,\big\vert\,  \mathbf{y}^\circ(t_1),\dots,\mathbf{y}^\circ(t_{n-1}) \right)d\mathbf{x}(t_{n-1}).
\]
One then assigns to a sample at position $\mathbf{x}(t_n)$ the importance weight
$p(\mathbf{y}^\circ(t_n)\,|\,\mathbf{x}(t_n)),$ proportional to
\[
\frac{{p}\left(\mathbf{x}(t_n)\, \big\vert\,   \mathbf{y}^\circ(t_1),\dots,\mathbf{y}^\circ(t_{n}) \right)}
{{p}\left(\mathbf{x}(t_n)\,\vert  \mathbf{y}^\circ(t_1),\dots,\mathbf{y}^\circ(t_{n-1})\right)}
\]
(the proportionality follows from \eqref{filterrecur}).

At the next observation time these steps are repeated.  The sample
weights from the previous observation time will be multiplied by
weights corresponding to the new observation.  Over several
observation times this leads to highly degenerate weights and poor
statistical properties of the resulting estimator.  To avoid this
problem the standard particle filter includes a simple resampling
step.  The full procedure is summarized in the following algorithm
(see \cite{gordon93}):
\begin{alg}[Basic particle filter]\label{pf1}
\item[]
{\tt
\begin{enumerate}
\item Begin with $M$ weighted samples
  $\left\{\left(\mathbf{x}_j(t_{n-1}),W_j(t_{n-1})\right)\right\}_{j=1}^M$ of\linebreak
  ${p}( \mathbf{x}(t_{n-1})\, \vert\, \mathbf{y}^\circ(t_1),\dots,\mathbf{y}^\circ(t_{n-1})).$
\item
For each $j,$ evolve $\mathbf{x}_j(t_{n-1})$ from time $t_{n-1}$ to time $t_n$ to generate an\linebreak independent sample
$\widetilde{\mathbf{x}}_j(t_n)$ from  ${p}(\mathbf{x}(t_n)\, \vert\, \mathbf{x}_j(t_{n-1}) ).$
\item
Evaluate the weights,
\[
\widetilde W_j(t_n) = p(\mathbf{y}^\circ(t_n) \,|\,\widetilde{\mathbf{x}}_j(t_n))\, W_j(t_{n-1}).
\]
\item Resample the particles if necessary to obtain a weighted
  ensemble\linebreak $\left\{\left(\mathbf{x}_j(t_n),W_j(t_n)\right)\right\}_{j=1}^M$
approximating $p(\mathbf{x}(t_n)\,|\,\mathbf{y}^\circ(t_1),\dots,\mathbf{y}^\circ(t_n)).$
\end{enumerate}
}
\end{alg}

\section{The small noise regime}
\label{sec:small noise}

Assume for the moment that in the particle filter, one resamples at
each step so that $W_j(t_n) = 1/M.$ Notice that Step 2 is identical in
both Algorithms~\ref{ek1} and~\ref{pf1}.  At the end of Step 2 both
the ensemble Kalman filter and the particle filter have generated an
empirical density,
\begin{equation}\label{empir1}
\frac{1}{M}\sum_{j=1}^M \delta(\mathbf{x}(t_n) - \widetilde{\mathbf{x}}_j(t_n))
\end{equation}
approximating ${p}(\mathbf{x}(t_n)\, \vert \mathbf{y}^\circ(t_1),\dots,\mathbf{y}^\circ(t_{n-1})).$ In the ensemble
Kalman filter this empirical density is approximated by a Gaussian
density allowing the information from the observation at time $t_n$ to
be incorporated analytically.  The particle filter uses the importance
weights to transform \eqref{empir1} into a weighted empirical density
approximating ${p}(\mathbf{x}(t_n)\, \vert  \mathbf{y}^\circ(t_1),\dots,\mathbf{y}^\circ(t_{n})).$

The difficulty with both methods is that the empirical density
\eqref{empir1} provides a very poor approximation of the tails of the
density ${p}(\mathbf{x}(t_n)\,\vert \mathbf{y}^\circ(t_1),\dots,\mathbf{y}^\circ(t_{n-1})).$ Of course most
observations (i.e. values of $\mathbf{y}^\circ(t_n)$) do not correspond to tail events
of ${p}(\mathbf{x}(t_n)\,\vert \mathbf{y}^\circ(t_1),\dots,\mathbf{y}^\circ(t_{n-1})).$ However, in many cases
(e.g. the Kuroshio), these tail events are precisely the most
interesting features of the system and the events that one is
primarily interested in capturing. 

The practical consequence of these tail events is easily understood by
considering the behavior of a particle filter on the Kuroshio.
Suppose that between two observations the hidden signal makes a
transition from one meander to another.  If at the time of the first
observation, all samples are in the small meander, then at best only a
few trajectories will make the transition to the large meander.
Therefore at the time of the next observation, when new likelihood
weights are calculated, almost all of the samples will receive
negligible weight, resulting in an estimator with very low accuracy.
As we will see in the next section the way to overcome this problem is
to use rare event sampling methods to
bias the evolution of samples toward regions of space where the
observation weights are large.  In order to introduce these methods,
let us first formalize the discussion on rare events: indeed many
methods have been proposed in the past to deal with these events (see
\cite{defreitas05, chorin2009_filter1, chorin2009_filter2}) but the
issues they pose do not seem to have been identified precisely.

To this end consider the situation in which the dynamics are governed
by the stochastic differential equation
\begin{align}
  &d{\mathbf{x}}(t) = F(\mathbf{x}(t)) 
  + \sqrt{\epsilon}\,\sigma(\mathbf{x}(t))\,d{B}(t)\notag\\
  &\mathbf{x}(0) = x_0 \label{eq:sde1}
\end{align}
where $B$ is Brownian motion, $F$ represents the deterministic components of the physical model, and $\sigma$ determines the size and covariance structure of noise arrising from unresolved physical features of the system or from model error.   Assume also that the
observations admit a conditional density of the form
\begin{equation}
p(\mathbf{y}^\circ(t_n)\,|\,\mathbf{x}(t_n)) \propto e^{-\frac{1}{\epsilon}g(\mathbf{y}^\circ(t_n),\mathbf{x}(t_n))}.
\end{equation}
The case of Gaussian $\xi_n$ variables  with covariance matrix $\epsilon \mathbf{R}$
corresponds to the choice
\begin{equation}
g(\mathbf{y}^\circ(t_n),\mathbf{x}(t_n)) = \frac{1}{2} (\mathbf{y}^\circ(t_n)- H(\mathbf{x}(t_n)))^\text{\tiny T} \mathbf{R}^{-1} (\mathbf{y}^\circ(t_n)- H(\mathbf{x}(t_n))).
\end{equation}
In these equations $\epsilon$ is a small parameter.  In this setup both
the stochastic forcing in the dynamics and the observation noise are
small.  {In most practical contexts $\epsilon$ will not explicity appear in the dynamics model  or in the observation model.   We stress that explicit knowlege of $\epsilon$ is not required to implement  the methods introduced below.  The factor of $\epsilon$ is introduced purely for the purpose of a formal analysis that will cleanly isolate the relevant performance issues.}

To make the case for rare event simulation tools in data assimilation
problems we will focus on the particle filter. However, as our
numerical tests will highlight, the difficulties presented by rare
events are shared by the Kalman filter family of methods as well.
Consider the assimilation of a single observation $\mathbf{y}^\circ(T)=\mathbf{y}^\circ$ at time
$T$ starting from initial condition $\mathbf{x}(0) = x_0.$ In particular
consider the weights $\widetilde W_j$ generated in Step 3 of Algorithm
\ref{pf1}.  When these weights are constant the samples $\widetilde{
\mathbf{x}}_j(T)$ generated in Step 2 are exactly samples from the posterior
density ${p}(\mathbf{x}(T)\,\vert \mathbf{y}^\circ).$ The variation in the weights is a
measure of the distance of the empirical density \eqref{empir1} from
the posterior density.  It is natural then to consider the behavior of
the relative standard deviation of the weights,
\begin{equation}\label{relerr}
\rho = \frac{\sqrt{\mathbf{E}\left[ \left( \widetilde W_j - \mathbf{E}\left[\widetilde
W_j\,\big|\,\mathbf{y}^\circ\right]\right)^2\,\big|\,\mathbf{y}^\circ\right]}}{\mathbf{E}\left[\widetilde
W_j\,\big|\,\mathbf{y}^\circ\right]}.
\end{equation}
 The weights at
observation time $T$ are identically distributed so the subscript
$j$ in the previous display can be ignored.  Expression \eqref{relerr}
can be rewritten as
\begin{equation}\label{relerr2}
\rho = \sqrt{R - 1}
\end{equation}
where
\begin{equation}\label{R}
R = \frac{\mathbf{E}\left[ \widetilde W_j^2\right]}{\mathbf{E}\left[\widetilde W_j\right]^2}
= \frac{\mathbf{E}\left[ e^{-\frac{2}{\epsilon}g(\mathbf{y}^\circ,\mathbf{x}(T))}\right]}{
\mathbf{E}\left[ e^{-\frac{1}{\epsilon}g(\mathbf{y}^\circ,\mathbf{x}(T))}\right]^2.}
\end{equation}
Note that $R\ge1$ since the expectation of the square of a random
variable is always greater than the square of its expectation.  If $R=1$ the
samples are drawn exactly from the posterior.  By dividing the number of samples by
$R$ one obtains the so called effective sample size, a rule of thumb
giving the number of unweighted samples that would be equivalent (in
terms of statistical accuracy) to an ensemble of weighted samples (see
for example \cite{defreitas05}).

Let us estimate $R$ in the present context. The Laplace Principle (see
e.g. \cite{Dembo:1993p6678,dupuisellisbook,Varadhan:1985p6658})
indicates that, under suitable assumptions,
\begin{equation}\label{eq:lap1}
 \lim_{\epsilon\rightarrow 0}-\log \mathbf{E}\left[e^{-\frac{1}{\epsilon}g\left(\mathbf{y}^\circ, \mathbf{x}(T)\right)}\right] = \gamma_1
  \end{equation}
which can alternatively be written
\begin{equation}\label{eq:lap1}
  \mathbf{E}\left[e^{-\frac{1}{\epsilon}g\left(\mathbf{y}^\circ, \mathbf{x}(T)\right)}\right]\asymp
e^{-\gamma_1/\epsilon}
  \end{equation}
  where two functions of $\epsilon$ are asymptotically equivalent
  ($\asymp$) if the ratio of their logarithms converges to 1 as
  $\epsilon\rightarrow 0.$ The constant $\gamma_1$ is given by the
  formula
  \begin{equation*}
  \gamma_1 =  \inf_{
	\varphi(0) = x_0} 
  \left\{ I\left(\varphi\right) +  g\left(\mathbf{y}^\circ,\varphi(T)\right)\right\}
\end{equation*}
where the infimum is taken over all absolutely continuous functions,
$\varphi,$ from $[0,T]$ into $\RR^d$ and
\begin{equation} \label{eq:action}
  I(\varphi) =  \int_0^{T} \frac{1}{2}\left(\dt \varphi(t)-
F(\varphi(t))\right)^\text{\tiny T}
\mathbf{Q}(\varphi(t))^{-1}\left(\dt \varphi(t)-
F(\varphi(t))\right)dt
\end{equation}
where \begin{equation} 
\mathbf{Q}(x) = \sigma(x)\sigma(x)^\text{\tiny T}
\end{equation}
is the covariance of the (incremental) model noise.
In our context this simply says that in the small noise regime the
posterior density is sharply peaked around the states that are most
likely (in the small noise limit) given the observations.  These most likely 
states are given by $\varphi(T)$ for $\varphi$ minimizing $I.$  The reader 
should note the similarity between the cost functional $I\left(\varphi\right) +  g\left(\mathbf{y}^\circ,\varphi(T)\right)$ and the cost functional
minimized in the weakly constrained variant of the 4DVar method (see \cite{Derber:1989:mwr}).

Applying the Laplace Principle one more time we obtain
\begin{equation}
\label{eq:lap2}
  \mathbf{E}\left[e^{-\frac{2}{\epsilon}g\left(\mathbf{y}^\circ, \mathbf{x}(T)\right)}\right] 
  \asymp e^{-\gamma_2/\epsilon}
  \end{equation}
  where
  \begin{equation*}
   \gamma_2 = \inf_{\varphi(0) = x_0} \left\{
    I\left(\varphi\right) 
    +  2g\left(\mathbf{y}^\circ,\varphi(T)\right)\right\}.
\end{equation*}
Notice that
\begin{equation}\label{rateineq}
\gamma_2 =  \inf_{
	\varphi(0) = x_0} \left\{
    I\left(\varphi\right) 
    +  2g\left(\mathbf{y}^\circ,\varphi(T)\right)\right\} \\
    \leq  \inf_{\varphi(0) = x_0} \left\{
   2 I\left(\varphi\right) 
    +  2g\left(\mathbf{y}^\circ,\varphi(T)\right)\right\} = 2\,\gamma_1.
\end{equation}
In most cases the inequality in \eqref{rateineq} is strict and
inserting (\ref{eq:lap1}) and (\ref{eq:lap2}) in
 \eqref{R} implies that 
$R$  increases exponentially  as $\epsilon\rightarrow 0$:   
\[
  R \asymp 
    e^{\frac{2\,\gamma_1-\gamma_2}{\epsilon}}.
\]
As pointed out above, our samples will have highest statistical quality when the weights 
are all equal (i.e. when $R=1$).  But as is clear from the last display, if $\gamma_2<2\gamma_1$ (and this is generically the case)
then the number of samples needed to maintain accuracy
increases exponentially as $\epsilon\rightarrow 0$. This is the primary
challenge posed by rare events.  Note that in our analysis we have treated the observations as fixed (and in particular independent of $\epsilon$).  
While this may at first seem strange, it is the appropriate choice given that our goal is to understand and correct the failure of standard filters
precisely at the moments they are faced with assimilating a (or a sequence of) very unlikely observations.  

Before closing this section we should remark that while the finite
time horizon setting (finite $T$) described here seems appropriate
for designing on-line filtering methods, it is not the appropriate
setting for studying the transition mechanism of the system itself.
This transition has a natural time scale (which probably has nothing
to do with the observation window $T$) on which it occurs.  Methods
for analyzing such events can be found in \cite{heymann08_gmam}.

\section{Importance sampling strategies}
\label{sec:importance}

The difficulty discussed in the previous section can be remedied
within the context of importance sampling as we now describe.  For a general discussion of importance sampling in the context
of particle filtering see, for example, \cite{defreitas05}.  The discussion will focus on our particular setting.  Recall
our assumption that the original dynamics of the system are governed
by the stochastic differential equation \eqref{eq:sde1}
\begin{align*}
  &d{ \mathbf{x}}(t) = F( \mathbf{x}(t))
  + \sqrt{\epsilon}\,\sigma(\mathbf{x}(t))\,d B(t)\notag\\
  &\mathbf{x}(0) = x_0
\end{align*}
(see e.g. \cite{karatzasandshreve}).
We now modify these dynamics by adding an additional forcing term $v(t,x)$ so
that instead of evolving \eqref{eq:sde1} we evolve
\begin{align}\label{Xv1}
&d{ \widehat {\mathbf{x}}}(t) = F( \widehat {\mathbf{x}}(t))+\sigma(\widehat{ \mathbf{x}}(t))\,v(t,\widehat{ \mathbf{x}}(t))
+ \sqrt{\epsilon}\,\sigma(\widehat{ \mathbf{x}}(t))\,d{B}(t) \\
&\widehat{ \mathbf{x}}(0) = x_0.\notag
\end{align}
The replacement of the underlying dynamics by a modified dynamics as in \eqref{Xv1} without statistical correction is sometimes referred to as  ``nudging''  (see \cite{Stauffer:1993:TellusA}).
To obtain unbiased statistics one much correct for this modification by computing an additional particle weight.
Girsanov's Theorem (see e.g. \cite{karatzasandshreve}) tells us that, for
reasonable choices of $v,$ the ratio of the probability of a path 
over the interval $[0,t]$ under the original dynamics (denoted informally by $P$)
to a path under the modified dynamics \eqref{Xv1} (denoted informally by $\widehat{P}$) is given by
\begin{equation*}
 {\frac{d   P}{d\widehat{P}  }} = e^{
-\frac{1}{\sqrt{\epsilon}}\int_0^{t} v(s, 
   \widehat{\mathbf{x}} (s))^\text{\tiny T} dB(s)\rangle  - \frac{1}{2\epsilon}\int_0^{t}
   v(s, \widehat{\mathbf{x}}(s))^\text{\tiny T}v(s, \widehat{\mathbf{x}}(s)) ds}.
\end{equation*}
Thus we can compute any expectation with respect to the original dynamics by
instead computing a weighted expectation over the modified dynamics.  Indeed, we have
\[
\mathbf{E}\left[ f(\mathbf{x})\right] = \int f(x) dP(x) 
= \int f(\widehat x)  {\frac{d  P }{d\widehat{P}  }}(\widehat x)
d\widehat{P}(\widehat x) = \mathbf{E}\left[ f(\widehat{\mathbf{x}}) {\frac{d   P }{d
\widehat{P}  }}(\widehat{\mathbf{x}})  \right].
\]
 In particular,  Steps {\tt ii} and {\tt iii} of Algorithm
\ref{pf1} can be replaced
by
{\tt
\begin{enumerate}
\item[ii']
For each $j,$ generate an independent sample $\widehat{\mathbf{x}}_j$ of the solution to
 the\linebreak modified stochastic differential equation
\begin{align*}
&d {\widehat{\mathbf{x}}}_j(t) = F( \widehat{\mathbf{x}}_j(t))+\sigma(\widehat{\mathbf{x}}_j(t))\,v(t,\widehat{\mathbf{x}}_j(t))
+ \sqrt{\epsilon}\,\sigma(\widehat{\mathbf{x}}_j(t))\,d{B}_j(t),\quad t\in [t_{n-1},t_n]\\
&\widehat{\mathbf{x}}_j(t_{n-1}) = \mathbf{x}_j(t_{n-1}).\notag
\end{align*}
\end{enumerate}
}
 and
{\tt
\begin{enumerate}
\item[iii']
Evaluate the weights,
\begin{equation*}
\widehat W_j = e^{-\frac{1}{\epsilon}g(\mathbf{y}^\circ(t_n), \widehat{\mathbf{x}}_j(t_n))}Z_j\, W_j(t_{n-1})
\end{equation*}
where
\begin{equation*}
Z_j = e^{ -\frac{1}{\sqrt{\epsilon}}\int_{t_{n-1}}^{t_n}  v(s, 
   \widehat{\mathbf{x}}_{j}(s))^\text{\tiny T} d{B}_j(s)  - \frac{1}{2\epsilon}\int_{t_{n-1}}^{t_n}
   v(s, \widehat{\mathbf{x}}_{j}(s))^\text{\tiny T}v(s, \widehat{\mathbf{x}}_{j}(s)) ds}.
\end{equation*}
\end{enumerate}
} The question then becomes how to choose $v$ in~\eqref{Xv1} so that
the ratio between the variance of the new weights $\widehat W_j$ and their
mean square remain small, and we avoid the problem discussed in the
previous section. This question is discussed next.

\subsection{The single observation case.}
Let us again briefly  assume that the process $\mathbf{x}$ is observed only once at time
$T$ and that $\mathbf{y}^\circ(T)=\mathbf{y}^\circ.$ 
Our goal is to choose the function $v$ so that the 
relative standard deviation of the weights in Step {\tt iii'} above is as small
as possible.
This is not a trivial task. 
 To
see what it entails, note that the relative standard deviation of the new
weights  is
\begin{equation*}
\rho = \sqrt{ R - 1}
\end{equation*}
where now (compare~\eqref{R})
\begin{align}\label{Rhat}
R &=  \frac{ \mathbf{E}\left[\left(e^{-\frac{1}{\epsilon}g(\mathbf{y}^\circ, \widehat{
\mathbf{x}}(T))}\frac{dP}{d\widehat{P}} \right)^2\right]}
{\mathbf{E}\left[e^{-\frac{1}{\epsilon}g(\mathbf{y}^\circ, \widehat{
\mathbf{x}}(T))}\frac{dP}{d\widehat{P}} \right]^2}\\
   &=
   \frac{ \mathbf{E}\left[e^{-\frac{2}{\epsilon}g(\mathbf{y}^\circ, \widehat{\mathbf{x}}(T))
-\frac{2}{\sqrt{\epsilon}}\int_0^{T}  v(s, 
   \widehat  { \mathbf{x}}(s))^\text{\tiny T} dB(s)\rangle  - \frac{1}{\epsilon}\int_0^{T}
    v(s, \widehat  { \mathbf{x}}(s))^\text{\tiny T} v(s, \widehat  { \mathbf{x}}(s))
ds}\right]}{\mathbf{E}\left[e^{-\frac{1}{\epsilon}g(\mathbf{y}^\circ,\mathbf{x}(T))}\right]^2}.
\end{align}

As we have explained before, to control the relative standard
deviation of the weights one must control $R$. Interestingly, there is
one choice of the biasing function $v$ which results in no statistical
error at all (i.e. such that $R=1$).  This choice will
turn out to be impractical.  However it will be useful to keep this ``gold
standard'' in mind as we consider more practical possibilities.  Specifically,
if we define
\[
\Phi(t,x) = \mathbf{E}_{t,x} \left[ 
e^{-\frac{1}{\epsilon}g(\mathbf{y}^\circ,\mathbf{x}(T))}\right].
\]
then it is possible to show that $R=1$ for every $\epsilon$ if we take
$v=v^\epsilon$ with
\begin{equation}\label{def:veps}
 v^\epsilon = \epsilon\frac{\sigma^\text{\tiny T}\,
D_x\Phi^{\epsilon}}{\Phi^{\epsilon}}
\end{equation}
where we have used $D_x$ to denote the gradient (viewed as a column vector).
 In other words the process
$\widehat{\mathbf{x}}$ in \eqref{Xv1} with $v = v^\epsilon$ samples exactly from the density
of $\mathbf{x}$ given the observation
at time $T.$ With this choice reweighting of the samples is unnecessary.  The
function $\Phi^{\epsilon}$ satisfies the backward Kolmogorov equation
\begin{equation}\label{eq:bke}
\partial_t\Phi^{\epsilon} +  F^\text{\tiny T} D_x\Phi^{\epsilon} +
\frac{\epsilon}{2}\, \text{tr}\,\mathbf{Q} D^2_x\Phi^{\epsilon}
= 0
\end{equation}
with terminal condition $\Phi(T,x) = e^{-\frac{1}{\epsilon}g(\mathbf{y}^\circ,x)}.$
In this expression $D^2_x \Phi^\epsilon$ is the matrix of second derivatives of $\Phi^\epsilon$
and
\begin{equation}
 \text{tr}\,A
 = \sum_{j=1}^m A_{jj}
\end{equation}
for any $m\times m$ matrix $A.$  We remind the reader that $\mathbf{Q} = \sigma \sigma^\text{\tiny T}.$
One could, therefore, attempt to discretize and solve the partial differential
equation \eqref{eq:bke} to 
approximate \eqref{def:veps}.  Unfortunately this strategy is impractical in
more than a few dimensions as it requires computations on a grid that is exponentially large in the dimension of $x.$
These ideas, however, can be modified and put to use in high dimensions.  
Notice that for each $\epsilon>0$ the function 
\begin{equation}\label{eq:Geps}
 G^{\epsilon} = -\epsilon \log \Phi^{\epsilon}
\end{equation}
 (in the differential equation liturature this is called a Hopf-Cole transformation) solves the second
order Hamilton--Jacobi equation
\begin{equation}\label{eq:hjeps}
-\partial_t G^\epsilon + \mathcal{H}(x,D_x G^\epsilon(t,x))
- \frac{\epsilon}{2}\,\text{tr}\,\mathbf{Q} D_x^2G^\epsilon = 0
\end{equation}
with terminal condition $G^{\epsilon}(T,x) = g(\mathbf{y}^\circ, x)$, where
\begin{equation}\label{eq:H}
 \mathcal{H}(x,p) = - p^\text{\tiny T} F(x)  + \frac{1}{2}\, p^\text{\tiny T} \mathbf{Q}p.
\end{equation}
In terms of $G^{\epsilon},$ the function $v^\epsilon$ in
\eqref{def:veps} can be written
\begin{equation}\label{def:veps2}
v^\epsilon = -\sigma^\text{\tiny T}\,D_xG^{\epsilon}.
\end{equation} 
Numerical solution of this equation is no more practical than solution of \eqref{eq:bke}.

We therefore continue our search for a more practical approximation of $v^\epsilon$ by replacing $G^{\epsilon}$ by its zero
viscosity approximation $G,$ i.e.
by the viscosity solution to the first order Hamilton--Jacobi equation,
\begin{equation}\label{eq:hj1}
-\partial_t G + \mathcal{H}(x,D_xG(t,x))=0
\end{equation}
with terminal condition $G(T,x)=g(\mathbf{y}^\circ,x),$ and to use the function
\begin{equation}\label{def:v0}
v^0 = -\sigma^\text{\tiny T}\, D_xG
\end{equation}
 in place of the choice in \eqref{def:veps}.

Direct solution of equation \eqref{eq:hj1} is, again, not practical in more than
a few dimensions.  However, under mild assumptions (see
\cite{Bardi:1997p5947,Fleming:2006p3998}), the solution of this
equation has an optimal control representation
  \begin{equation}\label{Gcont}
  G(t,x) = \inf_{\varphi(t) = x} 
  \left\{ I_{t,T}\left(\varphi\right) +  g\left(\mathbf{y}^\circ,
\varphi(T)\right)\right\}
\end{equation}
where
the infimum is taken over all absolutely continuous functions on $[t,t_1]$ and 
\begin{equation*}
  I_{t,T}(\varphi) =  \int_t^{T} \frac{1}{2}\left(\dt \varphi(s)- F(\varphi(s))\right)^\text{\tiny T}
(\mathbf{Q}(\varphi(s)))^{-1}\left(\dt \varphi(s)- F(\varphi(s))\right)
ds.
\end{equation*}
Notice that the constant $\gamma_1$ in formula \eqref{eq:lap1} is given by
$\gamma_1 = G(0,x_0).$

 Assume that at any point $(t,x)$ there is  one minimal trajectory in
\eqref{Gcont}, i.e.
 a trajectory $\widehat \varphi_{t,x}$ such that 
\begin{equation}\label{eq:optcontrol2}
 G(t,x) = I_{t,T}(\widehat\varphi_{t,x})+ g\left(\mathbf{y}^\circ, \widehat
\varphi_{t,x}(T)\right).
\end{equation}
Such a function  $\widehat \varphi_{t,x}$ will be called the optimal control
trajectory at $(t,x).$ It can be shown
that 
\begin{equation}\label{drifteq}
v^0(t,x) = \sigma^{-1}(x)\left(\dt{\widehat \varphi}_{t,x}(t) - F(x)\right).
\end{equation}

{F}rom expression \eqref{drifteq} it is clear that the choice
\eqref{def:v0} requires that one solve the variational problem
\eqref{Gcont} at each point along the path of $\widehat{\mathbf{x}}$ to find an
optimal control trajectory at $(s,\widehat{\mathbf{x}}(s))$ for each $s\in [0,T].$
Clearly, for problems in high dimensions solving \eqref{Gcont} is not
a trivial task and carrying this out ``on-the-fly'' could impose a
significant computational burden. In fact, the cost of generating each
sample trajectory of \eqref{Xv1} with this choice of $v$ is quadratic
in $T.$ However, when sampling sufficiently rare events this cost is
more than made up for by the favorable statistical properties of the
estimator corresponding to \eqref{def:v0}.  The results in
\cite{eveweare1} imply that for $v^0$ in \eqref{def:v0},
\[
\lim_{\epsilon\rightarrow 0}  R = 1.
\]
This should be contrasted with the case of the standard particle
filter for which we recall that $R$ grows exponentially with
$\epsilon^{-1}.$ Notwithstanding the results in \cite{eveweare1}, for
the extremely high dimensional models common in climate and weather
prediction it seems necessary to search for approximations which,
while they will not match the statistical performance of the choice in
\eqref{def:v0}, are less computationally burdensome.

In the following we suggest two possible choices for $v$ that in our
numerical experiments seem to be inexpensive and effective.  Both
methods require solving the optimization problem \eqref{Gcont} at
least once.  The optimization problem \eqref{Gcont} is highly
structured and in our numerical experiments the cost of the each
solution of \eqref{Gcont} is roughly ten times the cost of simulating
a single trajectory of the original system \eqref{eq:sde1}.  In fact
\eqref{Gcont} is very similar (particularly after discretization as in Appendix C) to the variational problem at the heart
of the weakly constrained 4DVar algorithm (see \cite{Derber:1989:mwr} and the references
therein) and it is likely that algorithms already in use within the
data assimilation community can be leveraged.

One natural further approximation of the optimal choice $v^\epsilon$ is obtained
by choosing a parameter $0<\tau\leq T$ and, for $k\leq T/\tau,$ solving
\begin{align}\label{Xv2}
&d {\widehat{\mathbf{x}}}(t) = F( \widehat{\mathbf{x}}(t))+\sigma(\widehat{\mathbf{x}}(t))\,v(t)
+ \sqrt{\epsilon}\,\sigma(\widehat{\mathbf{x}}(t))\,d{B}(t) \notag\\
&v(t) = v^0(t,\widehat\varphi_{k\tau,\widehat{\mathbf{x}}(k\tau)})\qquad \text{for}\qquad  t\in
[k\tau,(k+1)\tau].
\end{align}
Over each interval of length $\tau$ we have
simply replaced the spatial variable in $v^0$ by values along the optimal
trajectory $\widehat \varphi$ 
starting  from the position of the sample $\widehat{\mathbf{x}}$ at the end of the previous
length $\tau$ interval. 
Note that  on $[k\tau,(k+1)\tau]$  we can write
\[
v(t) = \sigma^{-1}(\widehat\varphi_{k\tau,\widehat{\mathbf{x}}(k\tau)}(t))\left( \dt{\widehat
\varphi}_{k\tau,\widehat{\mathbf{x}}(k\tau)}(t) - F( \widehat\varphi_{k\tau,\widehat{
\mathbf{x}}(k\tau)}(t))\right).
\]
For example, the choice $\tau=T$ yields
\begin{equation}\label{ct}
v(t) = v^0(t,\widehat\varphi_{0,x}(t)) =  \sigma^{-1}(\widehat\varphi_{0,x}(t))\left(
\dt{\widehat \varphi}_{0,x}(t) - F( \widehat\varphi_{0,x}(t))\right).
\end{equation}
As explained in \cite{eveweare1} the choice in \eqref{ct} can, in some cases, lead to disastrous performance and the exceptional statistical properties established there only hold for $v^0$ itself.  That said, on many problems approximations of $v^0$ such as the one in \eqref{ct} are highly effective.

\subsection{Some simple recursive schemes for the multiple observation case}
 
The picture that we have sketched changes slightly when assimilating
multiple sequential observations.  In general no recursive algorithm
can completely avoid poor performance as $\epsilon\rightarrow 0.$ To
see this suppose that between each observation in Algorithm \ref{pf1}
(with Steps {\tt ii'} and {\tt iii'}) we use the biasing function
$v^\epsilon.$ In the single observation case the choice $v^\epsilon$
led to perfect sampling.  In the multiple observation case however,
after assimilating the first observation we are forced to use some
approximation of the posterior density at time $t_1$ (an empirical
approximation in the case of Algorithm \ref{pf1}).  Unfortunately the
states at time $t_1$ that are most consistent with the observation at
time $t_2$ may be far into tail of the time $t_1$ posterior density.
Resolution in this tail will be very poor leading to large errors as
$\epsilon$ vanishes.

One solution to this problem is to go back to time $t_0$ and choose a
function $v$ that incorporates the observations at both $t_1$ and
$t_2.$ In principle the states at time $t_1$ that are most relevant to
any future observation (not just at $t_2$) may lie in the tail of the
time $t_1$ posterior density and one might have to ``backtrack'' many
observational steps each time a new observation arrives.  Though
potentially costly this backtracking is fairly easy to implement.
However, our numerical results indicate that significant statistical
improvements are possible without backtracking and we will not pursue
the topic further here.

The first algorithm that we suggest is a simple adaptation of the
biasing function in \eqref{Xv2} to the particle filtering framework:
\begin{alg}\label{pf3}
\item[]
{\tt
\begin{enumerate}
\item
Begin with $M$ unweighted samples $\{\mathbf{x}_j(t_{n-1})\}_{j=1}^M$ approximately drawn
from\linebreak
 ${p}( \mathbf{x}(t_{n-1})\,\vert\, \mathbf{y}^\circ(t_1),\dots,\mathbf{y}^\circ(t_{n-1})).$
\item
For each $j$
generate an independent sample $\widehat{\mathbf{x}}_j$ of the solution to \eqref{Xv2}
on\linebreak $[t_{n-1},t_n]$
starting from $\widehat{\mathbf{x}}_j(t_{n-1}).$  Label the corresponding trajectory of
$v$\linebreak in \eqref{Xv2}
by $v_j(t),$ $t\in[t_{n-1},t_n].$
\item
Evaluate the weights,
\begin{equation*}
\widehat W_j = e^{-\frac{1}{\epsilon}g(\mathbf{y}^\circ(t_n), \widehat{\mathbf{x}}_j(t_n))}Z_j\, W_j(t_{n-1})
\end{equation*}
where
\begin{equation*}
Z_j = e^{ -\frac{1}{\sqrt{\epsilon}}\int_{t_{n-1}}^{t_n}  v_j(s)^\text{\tiny T}
d{B}_j(s)  - \frac{1}{2\epsilon}\int_{t_{n-1}}^{t_n}
   v_j(s)^\text{\tiny T} v_j(s)ds}
\end{equation*}
\item Resample the particles if necessary to obtain a weighted
ensemble\linebreak 
$\left\{\left(\mathbf{x}_j(t_n),W_j(t_n)\right)\right\}_1^M$ approximating
$p(\mathbf{x}(t_n)\,|\,\mathbf{y}^\circ(t_1),\dots,\mathbf{y}^\circ(t_n)).$
\end{enumerate}
}
\end{alg}

While fairly simple and, in our tests, effective, Algorithm \ref{pf3}
still requires that one solve the optimization problem \eqref{Gcont}
once for each particle at each observation.  In practice this may not
be necessary.  The following {\it ad-hoc} algorithm solves the
optimization problem \eqref{Gcont} only once at each step.  The reader
should note the similiarity with the ensemble Kalman filter, though as
we will see, on some problems Algorithm \ref{ek2} has significant
advantages.
\begin{alg}
\label{ek2}
\item[]
{\tt \begin{enumerate}
\item
Begin with $M$ weighted samples
$\{\left(\mathbf{x}_j(t_{n-1}),W_j(t_{n-1})\right)\}_{j=1}^M$ approximating\linebreak
${p}( \mathbf{x}(t_{n-1})\,\vert\, \mathbf{y}^\circ(t_1),\dots,\mathbf{y}^\circ(t_{n-1})).$
\item
Determine  the sample mean $\mathbf{x}^a_{n-1}$ and covariance $\mathbf{P}^a_{n-1}$ of
the $\{\mathbf{x}_j(t_{n-1})\}_{j=1}^M.$
\item
Find the solution $\widehat \varphi$ of the optimization problem
\begin{multline*}
\inf_{\varphi} 
 \big\{ I_{t_{n-1},t_n}(\varphi) + 
g\left(\mathbf{y}^\circ(t_n), \varphi(t_n)\right)\\+ \frac{1}{2}
\left(\varphi(t_{n-1})-\mathbf{x}^a_{n-1}\right)^\text{\tiny T} (\mathbf{P}^a_{n-1})^{-1}\left(\varphi(t_{n-1})-\mathbf{x}^a_{n-1}\right) \big\}
  \end{multline*}
 \item
For each $j,$ generate an independent sample $\widehat{\mathbf{x}}_j$ of the solution to
\begin{equation*}
d {\widehat{\mathbf{x}}}_j(t) = F( \widehat{\mathbf{x}}_j(t))+\sigma(\widehat{\mathbf{x}}_j(t))\,v(t)
+ \sqrt{\epsilon}\,\sigma(\widehat{\mathbf{x}}_j(t))\,d{B}(t),\quad t\in [t_{n-1},t_n]
\end{equation*}
where $\widehat{\mathbf{x}}_j(t_{n-1})$ is a Gaussian random vector with mean
$\widehat\varphi(t_{n-1})$ and\linebreak covariance $\mathbf{P}^a_{n-1}.$
Here
\[
v(t) = \sigma^{-1}(\widehat\varphi(t))\left( \dt{\widehat \varphi}(t) - b(
\widehat\varphi(t))\right).
\]
\item
Compute the mean $\mathbf{x}^a_n$ and covariance $\mathbf{P}^a_n$ by
\[
\mathbf{x}^a_n = \frac{\sum_{j=1}^M \widehat{\mathbf{x}}_j(t_n) W_j(t_n)}{\sum_{j=1}^M  W_j(t_n)}
\quad\text{and}\quad \mathbf{P}^a_n = \frac{\sum_{j=1}^M (\widehat{\mathbf{x}}_j(t_n) - \mathbf{x}^a_n)(\widehat{\mathbf{x}}_j(t_n) - \mathbf{x}^a_n)^\text{\tiny T}W_j(t_n)}{\sum_{j=1}^M  W_j(t_n)}
\]
where
\begin{multline*}
W_j(t_n) = \\
\frac{\exp\left(-\frac{1}{2}(\widehat{
\mathbf{x}}_j(t_{n-1})-\mathbf{x}^a_{n-1})^{\text{\tiny T}} ({\mathbf{P}^a_{n-1}})^{-1}
(\widehat{\mathbf{x}}_j(t_{n-1})-\mathbf{x}^a_{n-1})\right)}
{\exp\left(-\frac{1}{2}(\widehat{
\mathbf{x}}_j(t_{n-1})-\widehat\varphi(t_{n-1}))^{\text{\tiny T}} ({\mathbf{P}^a_{n-1}})^{-1}
(\widehat{\mathbf{x}}_j(t_{n-1})-\widehat\varphi(t_{n-1}))\right)}
 e^{-\frac{1}{\epsilon}g(\mathbf{y}^\circ(t_n), \widehat{\mathbf{x}}_j(t_n))}Z_j
\end{multline*}
and
\begin{equation*}
Z_j = e^{ -\frac{1}{\sqrt{\epsilon}}\int_{t_{n-1}}^{t_n}
v(s)^\text{\tiny T}d{B}_j(s)  - \frac{1}{2\epsilon}\int_{t_{n-1}}^{t_n}
   v(s)^\text{\tiny T} v(s) ds}.
\end{equation*}
\end{enumerate}
}
\end{alg}
Notice that Algorithm \ref{ek2} only requires the solution of one
optimization problem per observation time.  In our tests the cost of
Step {\tt iii} is only about 10 times the cost of the evolution of a
single trajectory in Step {\tt iv}, i.e. the cost of running Algorithm
\ref{ek2} is comparable to the cost of running Algorithm \ref{pf1}
with 10 additional particles.  Thus it seems that this algorithm is
only marginally more expensive than the standard particle filter.  The
reader should appreciate that while this algorithm does assume that
the posterior is Gaussian, it does not assume that the predictive density is
Gaussian.  This is the key to its advantage over the ensemble Kalman
filter.  For some problems this algorithm may benefit from the
addition of a particle clustering step and a separate solution of the
optimization problem in Step {\tt iii} for each cluster (with
different mean and covariance for each cluster).  Finally we mention
that while  Algorithm \ref{pf3}  employs a resampling
step, one can easily imagine variants that do not.  For example, one
can modify these algorithms to incorporate the sample transformations
in \cite{bickel2009_nleaf}.  In the next section we will test Algorithms \ref{pf3} 
and \ref{ek2} against Algorithms \ref{pf1} and \ref{ek1} on a simple model of the Kuroshio.

\section{Our test problem and results}
\label{sec:model}

The filtering algorithms described in the previous sections 
will be tested on a stochastic perturbation of the
barotropic vorticity equation introduced by \cite{chao84} to model the
Kuroshio:
\begin{equation}\label{spde}
{\partial_t}{\mathbf{x}} + \frac{\partial}{\partial x}(u{\mathbf{x}})+ \frac{\partial}{\partial
y}(v{\mathbf{x}}) 
+ f\left(\frac{f_x}{f}-\frac{r_x}{r}\right)u +
f\left(\frac{f_y}{f}-\frac{r_y}{r}\right)v
= \nu\Delta {\mathbf{x}}
+ \sigma \eta
\end{equation}
Here $\mathbf{x}(t,x,y)$ is the vorticity, $r(x,y)$ is the water depth,
$f(x,y)$ is the Coriolis parameter, $\nu$ is the horizontal eddy
diffusivity, $(u,v)$ are the velocities in the $x$ and $y$ directions,
respectively, and $\eta$ is a space-time white noise whose amplitude
is controlled by $\sigma$.  A detailed description of the model boundary 
conditions and of $r$ can be found in Appendix B.

While this model  is far from a state of the art geophysical
representation of the Kuroshio, it is able to reproduce the
large and small meanders of the current: in the derterministic model,
these states are basins of attraction of the system forced by the
Kyushu wedge and Izu ridge (see Figure \ref{domain}).  Both meanders
coexist only for certain inflow volume conditions.  For the
deterministic model ($\sigma=0$) Chao demonstrated that it is possible
to observe the transition between meanders by varying the inflow
condition.  The stochastic modification~\eqref{spde} of Chao's model
was introduced by one of the authors in \cite{wearephd} and
\cite{weare09} where it was used to test a path sampling based
filtering algorithm.

\begin{figure}[t] \noindent
\includegraphics[width=33pc,angle=0]{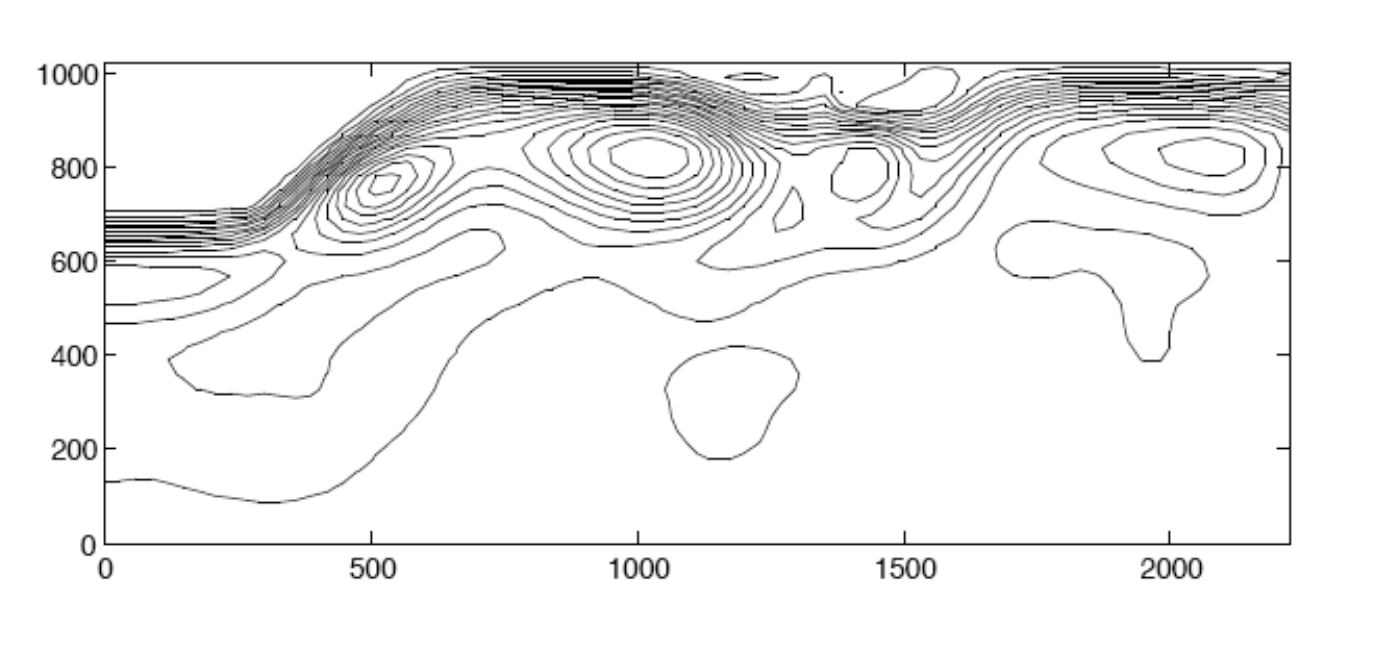}
\caption{ Small meander state of the model, \eqref{spde}.
The domain is as depicted in Figure \ref{domain}. }
\label{mysmallmeander}
\end{figure}

\begin{figure}[t] 
\noindent
\includegraphics[width=33pc,angle=0]{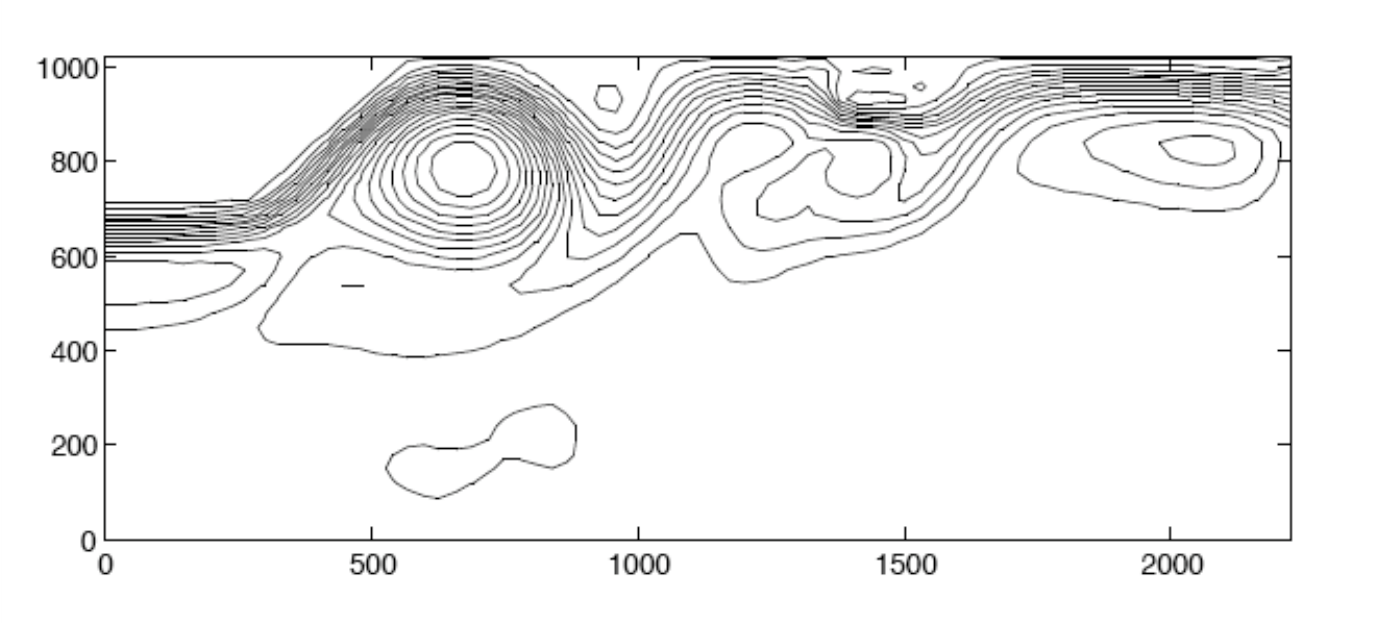}
\caption{
Large meander state of the model, \eqref{spde}.
The domain is as depicted in Figure \ref{domain}. }
\label{mylargemeander}
\end{figure}

In our calculations, we discretized \eqref{spde} using a uniform $30$
kilometer grid exactly as described in \cite{weare09,wearephd} and in
Appendix B.  The 2516 dimensional discrete time process so obtained
(which we also denote by $\mathbf{x}$) exhibits two meta-stable states that are
qualitatively similar to the small and large meanders of the actual
Kuroshio.  Figures \ref{mysmallmeander} and
\ref{mylargemeander} show typical states in both of these meanders.
They were found by varying the northern boundary condition as in
\cite{chao84}.  The noise parameter in \eqref{spde} ($\sigma$) was set
to 0 for the purposes of generating Figures \ref{mysmallmeander} and
\ref{mylargemeander}.

\begin{figure}[t] \noindent
\includegraphics[width=33pc,angle=0]{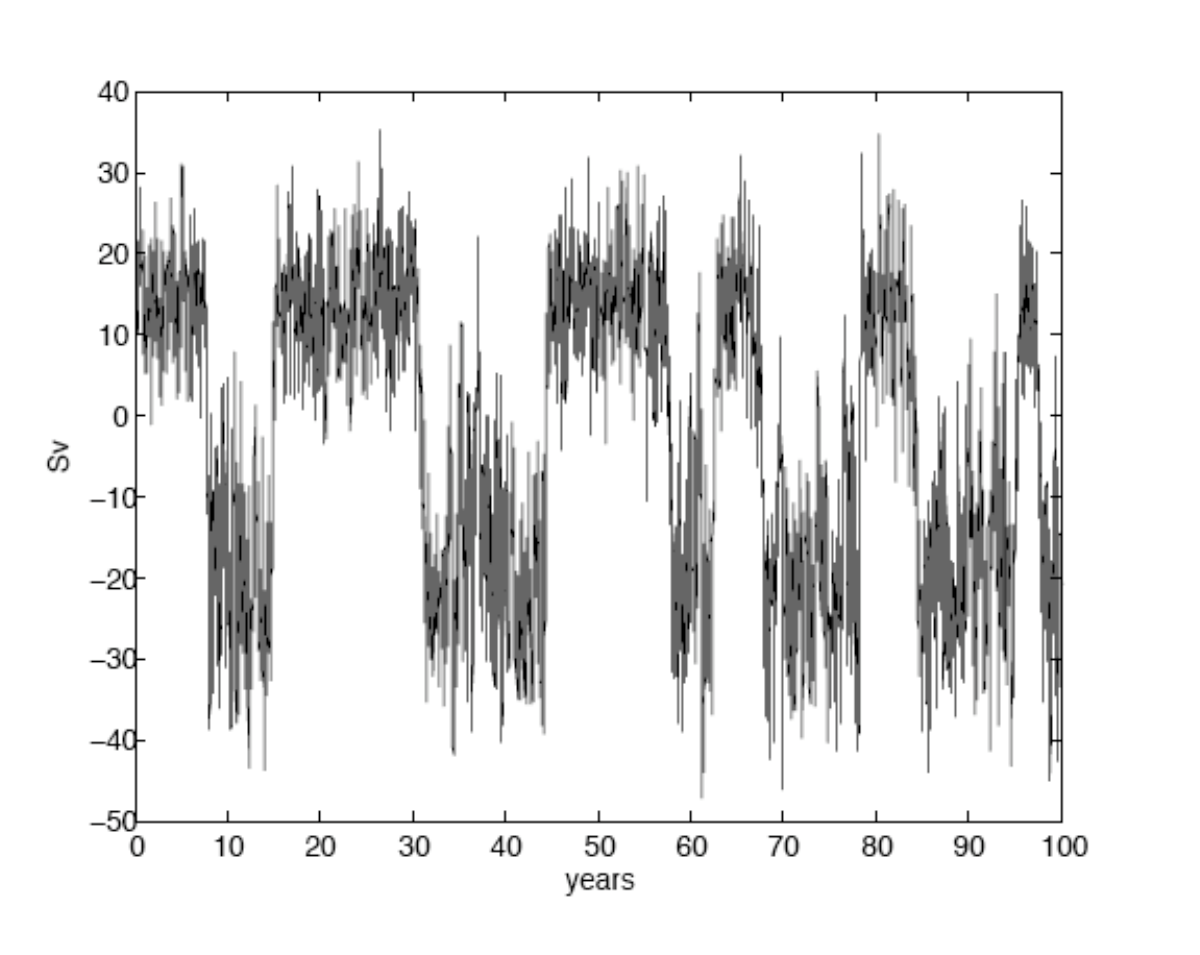}
\vspace{-24pt}
\caption{   Time series of approximation to $\psi(x^*,y^*)$ where
$(x^*,y^*)$ denotes the point in $D$ that is 990$km$ from the western
boundary and 860$km$ from the southern boundary.  Notice the transitions
between a metastable state near $15Sv$ and one near $-20Sv.$}
\label{obspoint}
\end{figure}

 Let $(x^*,y^*)$ denote the point in $D$ that is 990$km$ from the western
boundary and 860$km$ from the southern boundary.  This point is pictured in
Figure \ref{domain}. The bimodality of the system is evident in  Figure
\ref{obspoint} which shows a long trajectory 
of the system 
projected onto the variable $\psi(x^*,y^*).$  The state for which
$\psi(x^*,y^*)\approx 15 Sv$ roughly corresponds to the small meander and the
state
for which $\psi(x^*,y^*)\approx -20 Sv$ roughly corresponds to the large
meander (`$Sv$' stands for `Sverdrup' and  1 $Sv$ represents 
a volume transport of $10^{-3}\, km^3 s^{-1}$).
As can also be seen in Figure \ref{obspoint} the discrete stochastic system
tends to remain in each of its meanders for roughly 10 years.  Transitions
between the two meanders usually occur in a time 
span of a few months (though we again caution that this model is far from state
of the art).

Conditioned on the state $\mathbf{x}(t_n),$ the observation at time $t_n,$ $\mathbf{y}^\circ(t_n),$ is a
Gaussian random variable with mean $\psi(t_n)_{k^*,m^*}$ and standard deviation
1.92918.
Thus the vorticity process $\mathbf{x}(t)$ is observed through the value
of the corresponding volume transport process $\psi(t_n)$ at the
single point $(k^*,m^*).$ The observation times are separated by 2.63
days.   In our tests we assume that the initial state is known exactly to avoid initialization 
issues.

\subsection{Test results}\label{sec:results}

\begin{figure}[t]\noindent
\includegraphics[width=33pc,angle=0]{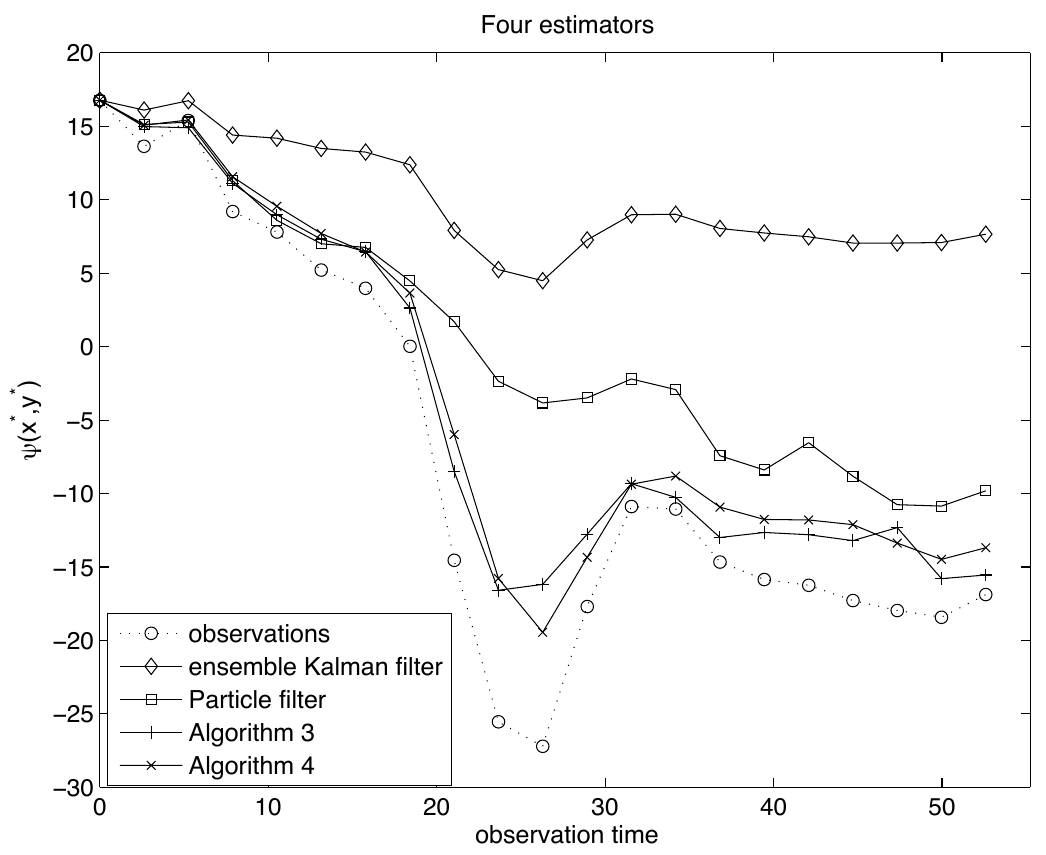}
\caption{Trajectory of estimate of  $\psi({ x}^*,{ y}^*)$
given observation for the four algorithms outlined in this paper.
$({x}^*,{ y}^*)$ is the point in $D$ that is 990$m$ from the western
boundary and 860$m$ from the southern boundary.  The particle filter, the
ensemble Kalman filter, and Algorithm \ref{ek2} are all run with 100 particles
while Algorithm \ref{pf3} is run with 
10 particles.  The estimates produced by Algorithms \ref{pf3} and \ref{ek2}
track the signal more closely than the ensemble Kalman filter and the particle
filter.}
\label{estimates}
\end{figure}

\begin{figure}[t] \noindent
\includegraphics[width=33pc,angle=0]{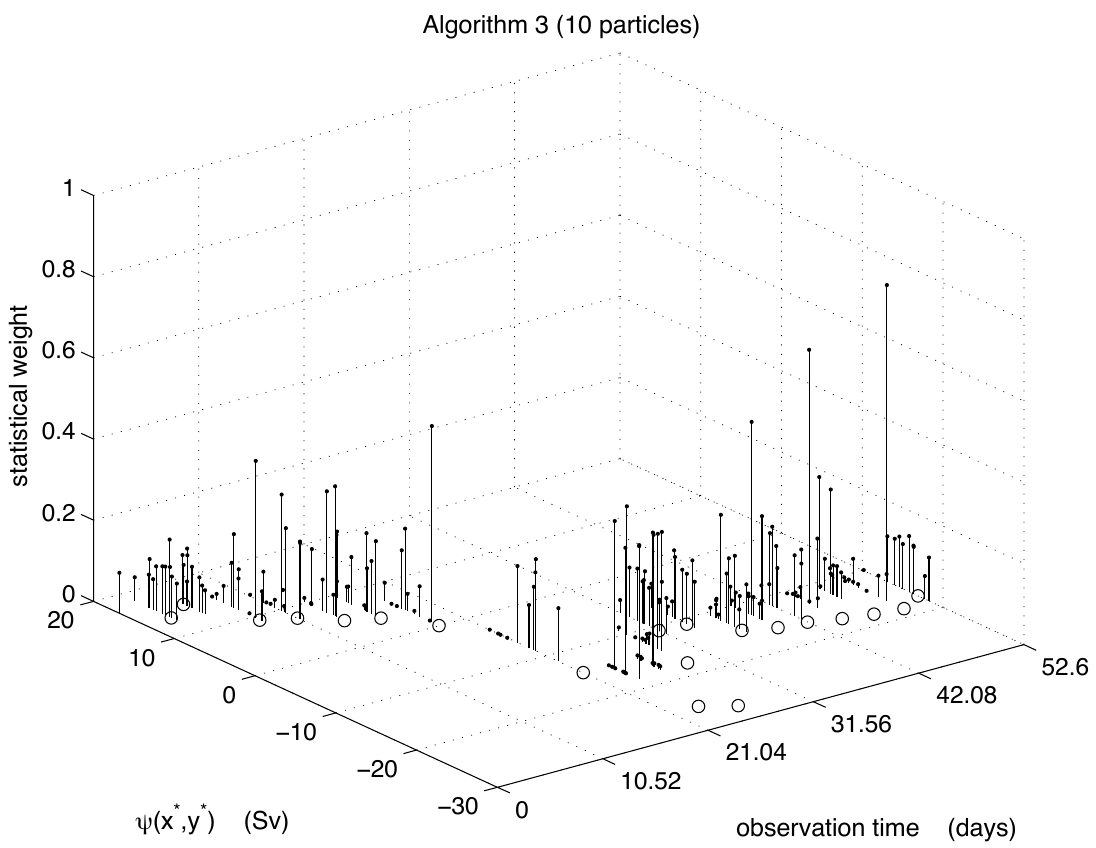}
\caption{ Plot of the values of $\psi({ x}^*,{ y}^*)$
corresponding to the samples $\widehat{\mathbf{x}}_j$ generated
in Step {\tt ii} of Algorithm \ref{pf3} (with 10 particles)
 at each observation time.  The $z$-coordinate of each point is the value
 of the weights (normalized) computed in Step {\tt iii} of the algorithm. Most
the 10 particles have weight comparable to the mean weight (in this case 0.1)
and the particle values
 of $\psi({ x}^*,{ y}^*)$ are clustered near the observation.} 
\label{dynstems}
\end{figure}

\begin{figure}[t]\noindent
\includegraphics[width=33pc,angle=0]{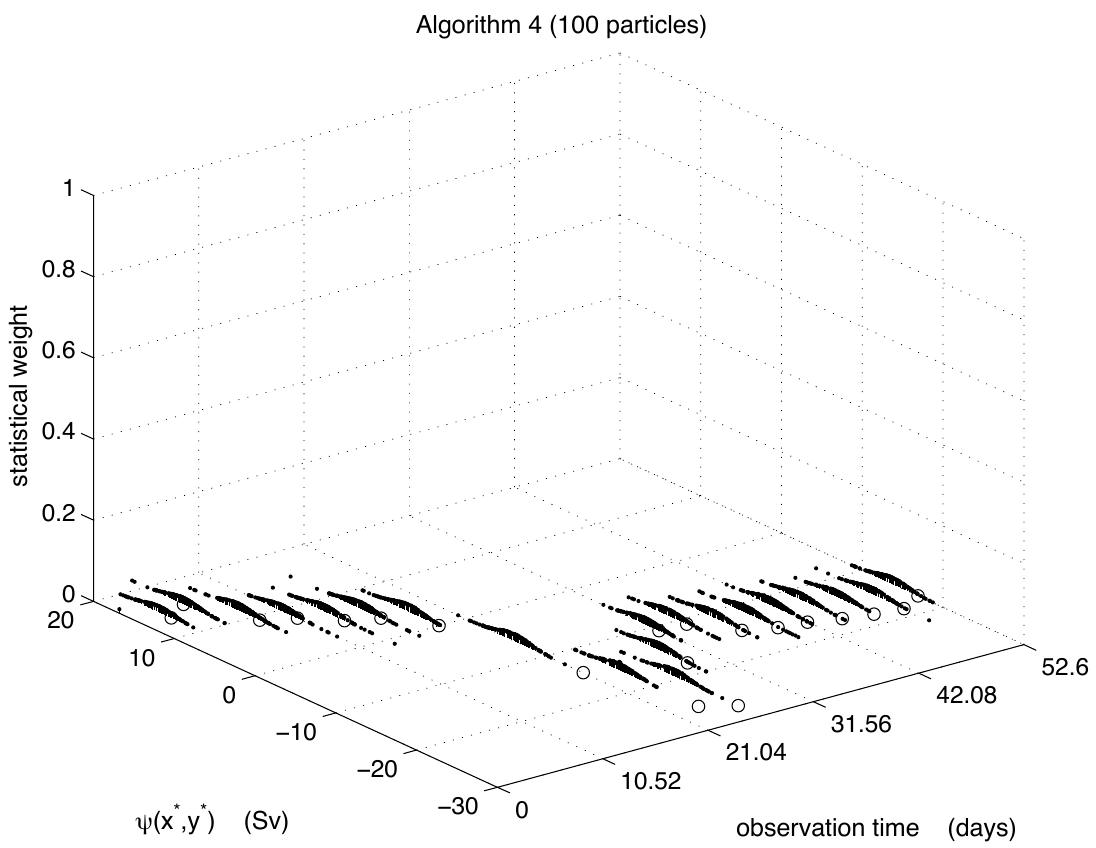}
\caption{ Plot of the values of $\psi({ x}^*,{ y}^*)$
corresponding to the samples $\widehat{\mathbf{x}}_j$ generated
in Step {\tt iv} of Algorithm \ref{ek2} (with 100 particles)
 at each observation time.  The $z$-coordinate of each point is the value
 of the weights (normalized) computed in Step {\tt v} of the algorithm.  All 100
particles have weight comparable to the mean weight (in this case 0.01)  and the
particle values
 of $\psi({ x}^*,{ y}^*)$ are clustered near the observation.}
\label{ot5stems}
\end{figure}

\begin{figure}[t] \noindent
\includegraphics[width=33pc,angle=0]{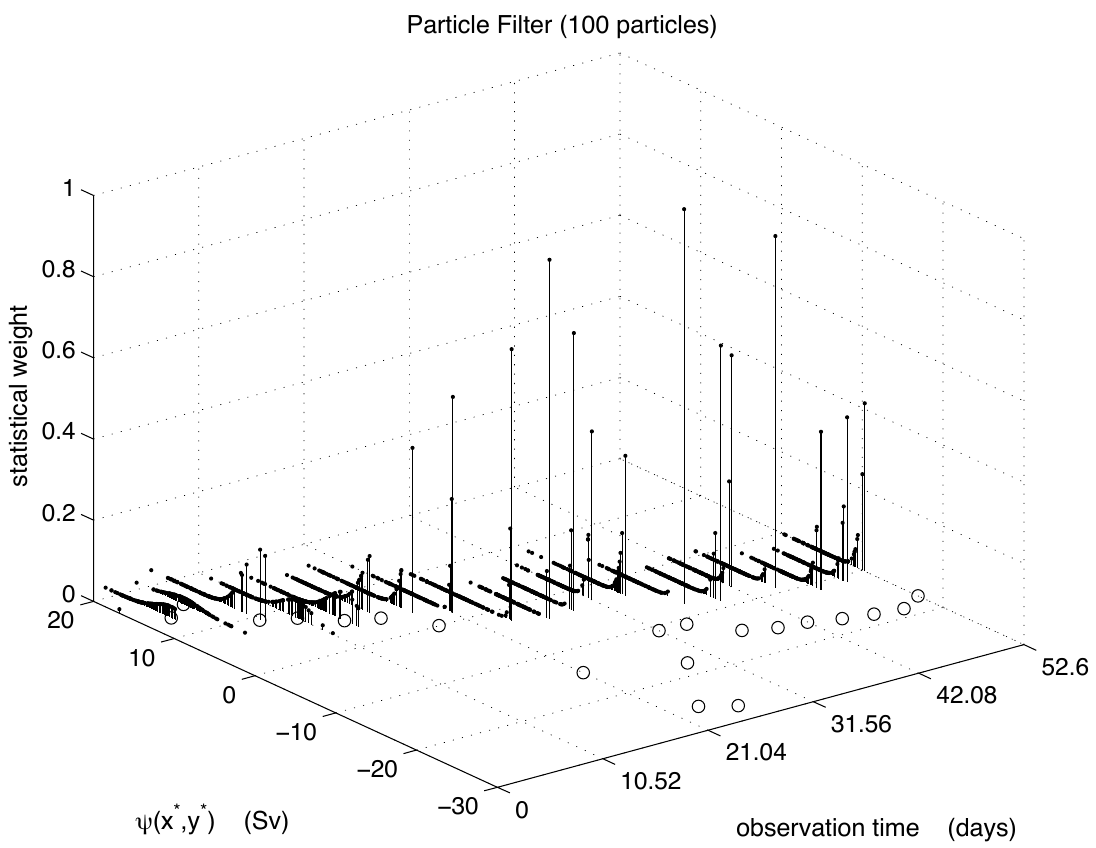}
\caption{Plot of the values of $\psi({ x}^*,{y}^*)$
corresponding to the samples $\widehat{\mathbf{x}}_j$ generated
in Step {\tt ii} of the standard particle filter (with 100 particles)
 at each observation time.  The $z$-coordinate of each point is the value
 of the weights (normalized) computed in Step {\tt iii} of the algorithm.  At
several steps there is only one particle with appreciable weight and that
particle's 
 value of $\psi({ x}^*,{ y}^*)$ is far from the observations compared to the
corresponding values generated by Algorithms~\ref{pf3} and~\ref{ek2}.}
\label{particlestems}
\end{figure}

\begin{figure}[t] \noindent
\includegraphics[width=33pc,angle=0]{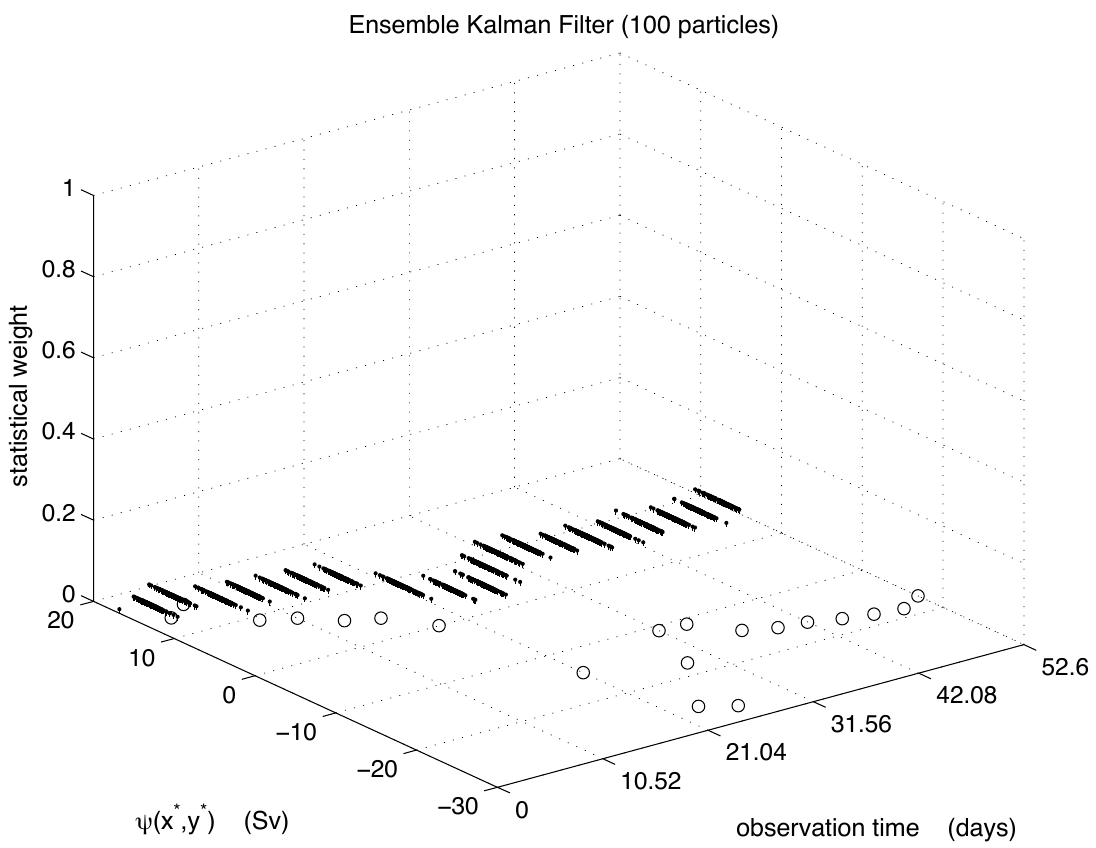}
\caption{Plot of the values of $\psi({ x}^*,{y}^*)$ corresponding to
  the samples $\widehat{\mathbf{x}}_j$ generated in Step {\tt ii} of the ensemble
  Kalman filter (with 100 particles) at each observation time.  The
  $z$-coordinate of each point is $0.01$ (the weights in this scheme
  are constant).  The particle values of $\psi({ x}^*,{ y}^*)$ are far
  from the observations compared to the corresponding values generated
  by Algorithms~\ref{pf3} and~\ref{ek2}.  }
\label{enkalstems}
\end{figure}

Using the setup described in the previous section we perform a simple
but revealing test of the ability of the four filters discussed in
this paper to track a single transition of the system from one meander
to another.  All Algorithms begin from a single initial state in the
small meander.  The observations are plotted in Figures
\ref{estimates} and \ref{dynstems}--\ref{enkalstems}, and correspond
to a segment of a long simulation of the system in which the system
transitions from the small meander to the large meander.  We do not
add noise to these observations.  The resulting estimates produced by
each of the four algorithms are displayed in Figure \ref{estimates}.
Evidently Algorithms \ref{pf3} and \ref{ek2} are able to properly
assimilate the change in the underlying state while the particle
filter and the ensemble Kalman filter do not perform as well.  The
ensemble Kalman filter completely misses the transition while the
particle filter seems to eventually capture it but remains
relatively far from the estimates produced by Algorithms~\ref{pf3}
and~\ref{ek2}.

While for a large enough number of samples the particle filter would
give the correct result, it is important to observe that the results
for the ensemble Kalman filter are fully converged.  No matter how
many particles are used in the ensemble Kalman filter it will not be
able to effectively track the transition.  This is because, as
discussed earlier,  the ensemble Kalman filter uses an extremely
poor approximation of the
tail of the predictive density, resulting in an
estimate that is nearly arbitrary in problems highly sensitive to tail
characteristics.  A similar phenomenon was observed for the extended Kalman filter in 
\cite{miller:1994:jas}.  The particular transition that we have used to
define the observations is more rapid than most of the transitions
that we observed in our long simulation of the system.  On a slower
transition the particle filter will show improved results. The
ensemble Kalman filter performed poorly in all the cases that we
examined. In our implementations of the ensemble Kalman filter and
Algorithm \ref{ek2} we make the assumption that  the posterior
covariance matrix ( $\mathbf{P}^a_n$) is diagonal.  In the ensemble Kalman filter we also
assume that the predictive covariance matrix ($\mathbf{P}_n$) is diagonal.

\begin{figure}[t]\noindent
\includegraphics[width=33pc,angle=0]{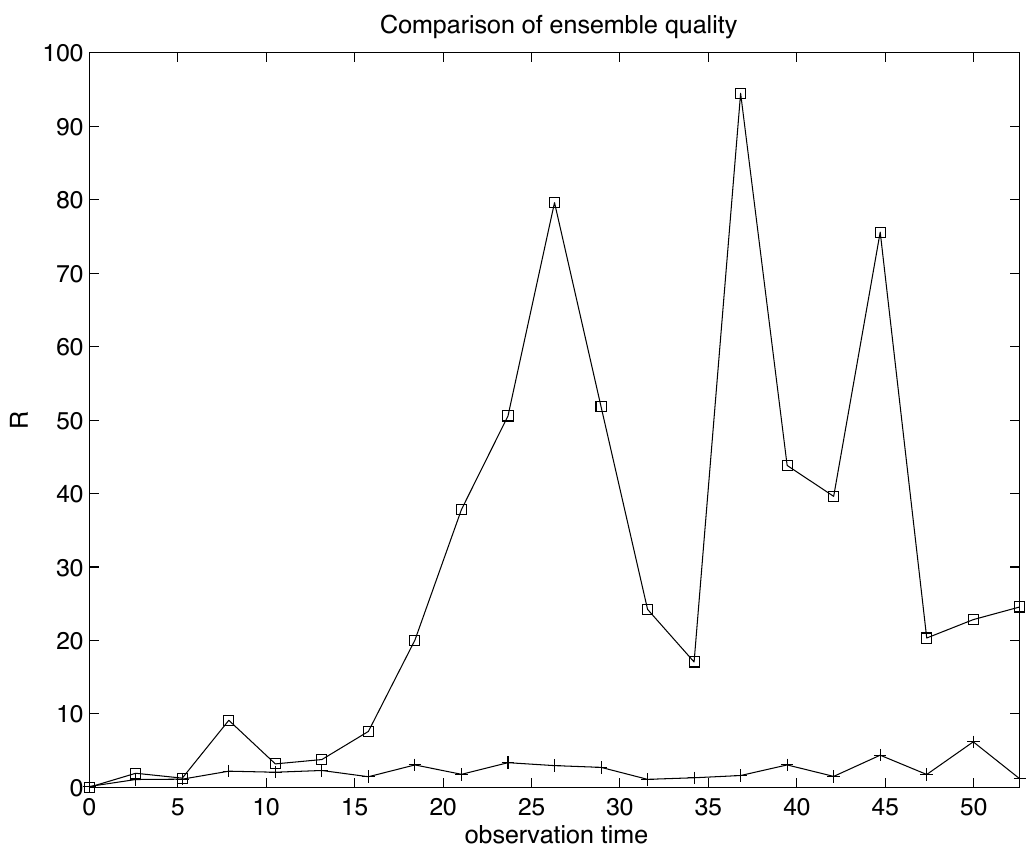}
\caption{ The values of $R$ defined in equations \eqref{R} (for the
  particle filter with 100 particles) and $\eqref{Rhat}$ (for
  Algorithm \ref{pf3} with 10 particles).  The effective sample size
  (see for example \cite{defreitas05}) for the two algorithms is found
  by dividing the number of particles by $R-1$. We plot the $R$ values
  only for the two statistically consistent schemes (Algorithm
  \ref{pf3} and the particle filter), that is the schemes that we know
  to converge to the correct density as the number of samples is
  increased.  This plot suggests that the ensemble generated by
  Algorithm \ref{pf3} is of much higher statistical quality than the
  ensemble generated by the particle filter in the sense that to
  achieve the same accuracy as Algorithm \ref{pf3} the particle filter
  would require as many as 100 times the number of samples.  It is not
  a meaningful statistic for approximate schemes such as Algorithm
  \ref{ek2} or the ensemble Kalman filter.}
\label{Rvals}
\end{figure}

Only two of the schemes tested (Algorithm \ref{pf3} and the particle
filter) are statistically consistent in the sense that they exactly
reproduce the posterior mean given a large enough sample size.  For
these two schemes we give the values of $R$ (defined in \eqref{R} and
\eqref{Rhat}, respectively) in Figure~\ref{Rvals}. Recall that by
dividing the number of samples by $R$ one obtains the
so-called effective sample size, a rule of thumb giving the number of
unweighted samples that would be equivalent (in terms of statistical
accuracy) to an ensemble of weighted samples (see for example
\cite{defreitas05}). The results suggest that the ensemble generated
by Algorithm \ref{pf3} is of much higher statistical quality than the
ensemble generated by the particle filter in the sense that to achieve
the same accuracy as Algorithm~\ref{pf3} the particle filter would
require as many as 100 times the number of samples. We do not compare
similar statistics for the two more approximate schemes (the ensemble
Kalman filter and Algorithm~\ref{ek2}).

We run the particle filter and the ensemble Kalman filter with 100
particles each ($M=100$).  Algorithm \ref{pf3} is run with only 10
particles ($M=10$).  Algorithm \ref{ek2} is run with 100 particles
($M=100$).  In Algorithm \ref{pf3} we set $\tau = 0.263$ days (see
expression \eqref{Xv2}).  With these choices, the particle filter, the
ensemble Kalman filter, and Algorithm \ref{ek2} have roughly equal
cost.  Algorithm \ref{pf3} is several times more expensive than the
others.  The optimizations in Algorithm \ref{pf3} and \ref{ek2}
typically converge to an acceptable level of accuracy in a handful of
iterations (using a total of 10 Jacobian multiplications or less).
For more details on how these optimizations are carried out see
Appendix C.  Alternative computational  approaches to similar optimization 
problems can be found in \cite{e:2004cpam} and \cite{eveweare1}.

In Figures \ref{dynstems}--\ref{enkalstems} we represent the empirical
densities produced by the four algorithms.  In each figure a vertical
line is drawn at each sample value of $\psi(x^*,y^*)$ at every
observation time.  The height of each line is the weight of that
sample.  One can clearly see that the ensembles generated by the
particle filter are dominated by a single particle at many of the
observation times while the other schemes produce more regular
weights.  Again we can see that the ensembles generated by the
particle filter and the ensemble Kalman filter are concentrated much
farther from the observations than the ensembles generated by
Algorithms \ref{pf3} and \ref{ek2}.  Of course proximity to the
observation is not a reliable measure of success in filtering
(proximity to the posterior density is our goal).  However, given that
both Algorithms \ref{pf3} and \ref{ek2} agree, we conclude that they
are accurate.

\section{Conclusion}
\label{sec:conclusion}

We have suggested that the failure of standard filtering algorithms
such as  ensemble Kalman filters and particle filters on certain
problems can be easily understood and analyzed in a regime in which the systems 
stochastic forcing (due to physics and model error) and the observational noise (due to measurement error) are both small.
As a first attempt at overcoming these issues we propose two schemes
based on rare event simulation tools.  The field of rare event
simulation has seen explosive growth over the last decade or so and
our investigation here is far from an exhaustive study of the
potential utility of rare event tools within data assimilation.
However, our results do strongly indicate that more
investigation in this direction is warranted.

In the context of the Kuroshio where standard methods fail, our new
methods inspired by rare event simulation techniques are very
effective.  Of course, the application of these methods is not limited
to the study of the Kuroshio.  Even restricting our attention
to geophysics, one can imagine important applications such as the
prediction of short and medium term extreme geophysical events like
tsunamis, hurricanes, and drought, as well as paleoclimate data
assimilation. We also remark that the hybrid nature of our schemes
suggest that they could be combined with existing large scale data
assimilation code employed in a black box fashion to design new robust
algorithms.

In our numerical test we have chosen to focus on a system with an
obvious, interesting tail feature (the meander transition).  While
problems of this kind provide stark evidence that current filtering
practice can be inadequate, it is important to point out that the same
failures can occur on problems with more mundane tail behaviors.
Indeed, over long periods of time one can expect nearly any underlying
(``hidden'') process to undergo rare excursions.  In exactly the way
that we have demonstrated, these excursions will cause standard
methods to loose track of the signal.

Finally we would like to point out that the methods proposed here are
 related to schemes proposed in \cite{chorin2009_filter1,
  chorin2009_filter2} and called ``implicit sampling'' methods.  In
those schemes one attempts to construct a map taking an easily sampled
random variable (usually a multidimensional Gaussian) to a random
variable distributed according to the posterior density.  The function
$v^\epsilon$ defined in \eqref{def:veps} is an exact solution to this
problem.  It can be used to map the trajectory of a Brownian motion
(an easily sampled random variable) to a random variable (the
corresponding $\widehat{\mathbf{x}}(T)$) drawn exactly from the posterior density.
In the low noise regime that we focus on, good statistical behavior
does not require that the mapping be exact.  For example, the
theoretical results in \cite{eveweare1} indicate that an asymptotic
(in the small noise limit) approximation of $v^\epsilon$ ($v^0$ in
\eqref{def:v0}) can yield strikingly good error behavior. The
importance sampling schemes at the core of Algorithms \ref{pf3} and \ref{ek2}
are
based on further approximations of $v^\epsilon$ and perform extremely well in
our tests.

\section*{Acknowledgments}
We would like to thank Professors Dorian Abbot, Alexandre Chorin, and Juan
Restrepo
for helpful conversations and the referees for their many detailed  suggestions.
  We would also like to thank Professor Chorin for
 suggesting the Kuroshio to JW as an interesting test system. 
  EVE was supported by ONR through award N00014-11-1-0345, by NSF through award DMS-0708140, and by DOE through award DE-SC0002618.
   The efforts of JW were supported by NSF through award DMS-1109731.

\appendix

\section{A derivation of $v^\epsilon.$}
Consider the setting in which there is a single observation $\mathbf{y}^\circ(T) = \mathbf{y}^\circ$ of the solution to
\begin{align*}
  &d{ \mathbf{x}}(t) = F( \mathbf{x}(t))
  + \sigma(\mathbf{x}(t))\,d B(t)\notag\\
  &\mathbf{x}(0) = x_0
\end{align*}
is collected at time $T.$   In the text it was claimed that if, in our importance sampling framework, we choose 
\begin{equation*}
 v = \frac{\sigma^\text{\tiny T}\,
D_x\Phi}{\Phi}
\end{equation*}
where
\[
\Phi(t,x) = \mathbf{E}_{t,x} \left[ 
e^{-g(\mathbf{y}^\circ,\mathbf{x}(T))}\right]
\]
 then $R=1$ where $R$ is defined in \eqref{Rhat}.  Here
\begin{align*}
  &d{\widehat{ \mathbf{x}}}(t) = F( \widehat{\mathbf{x}}(t))
+ \sigma(\widehat{\mathbf{x}}(t)) v(t,\widehat{\mathbf{x}}(t))
  + \sigma(\widehat{\mathbf{x}}(t))\,d B(t)\notag\\
  &\widehat{\mathbf{x}}(0) = x_0
\end{align*}
 We omit the various factors of $\epsilon$ which are irrelevant in this argument.  

If we can show that the wieghts 
\[
\widetilde W = e^{-g(\mathbf{y}^\circ,\widehat{\mathbf{x}}(T))}\exp\left(-\int_0^T v(t,\widehat{\mathbf{x}}(t))^\text{\tiny T} dB(t) - \frac{1}{2}\int_0^T v(t,\widehat{\mathbf{x}}(t))^\text{\tiny T}v(t,\widehat{\mathbf{x}}(t)) dt\right)
\]
are not random then it follows immediately that $R=1.$
To that end, first note that
\[
\widetilde W = \Phi(T,\widehat{\mathbf{x}}(T)) Z(T)
\]
where here
\[
Z(t) = \exp\left(-\int_0^t v(s,\widehat{\mathbf{x}}(s))^\text{\tiny T} dB(s) - \frac{1}{2}\int_0^t v(s,\widehat{\mathbf{x}}(s))^\text{\tiny T}v(s,\widehat{\mathbf{x}}(s)) ds\right).
\]
An application of It\^o's formula (see  e.g. \cite{karatzasandshreve}) to the product $\Phi(t,\widehat{\mathbf{x}}(t)) Z(t)$ (and omitting the arguments $t$ and $\widehat{\mathbf{x}}(t)$)
yields 
\begin{align*}
d{\left(\Phi Z\right)} &= Zd{\Phi} + \Phi d Z + Z v^\text{\tiny T}\sigma^\text{\tiny T} D_x  \Phi dt\\
&= Z\left( \partial_t \Phi dt + D_x \Phi^\text{\tiny T} d\widehat{\mathbf{x}} + \text{tr}\, \mathbf{Q} D^2_x\Phi \,dt \right) 
- Z \Phi  v^\text{\tiny T}dB -  Z v^\text{\tiny T}\sigma^\text{\tiny T} D_x  \Phi dt\\
& = Z\left( \partial_t \Phi +  D_x \Phi^\text{\tiny T} F
 + \text{tr}\, \mathbf{Q} D^2_x\Phi\right) dt + Z\left(D_x \Phi^\text{\tiny T} \sigma v -v^\text{\tiny T}\sigma^\text{\tiny T} D_x  \Phi  \right) dt\\
&\hspace{2cm} +  Z\left( D_x \Phi^\text{\tiny T} \sigma  - \Phi v^\text{\tiny T} \right) dB.
\end{align*}
Plugging in the choice of $v$ above we see that the right hand side in last display vanishes and therefore that
\[
\Phi(t,\widehat{\mathbf{x}}(t)) Z(t)
\]
is constant in $t.$
We conclude that
\[
\widetilde W = \Phi(T,\widehat{\mathbf{x}}(T)) Z(T)) = \Phi(0,x_0).
\]
Since the quantity on the right is not random the argument is complete.

\section{Model details and discretization}
\begin{figure}[t]
\noindent
\includegraphics[height=16pc, width=36pc,angle=0]{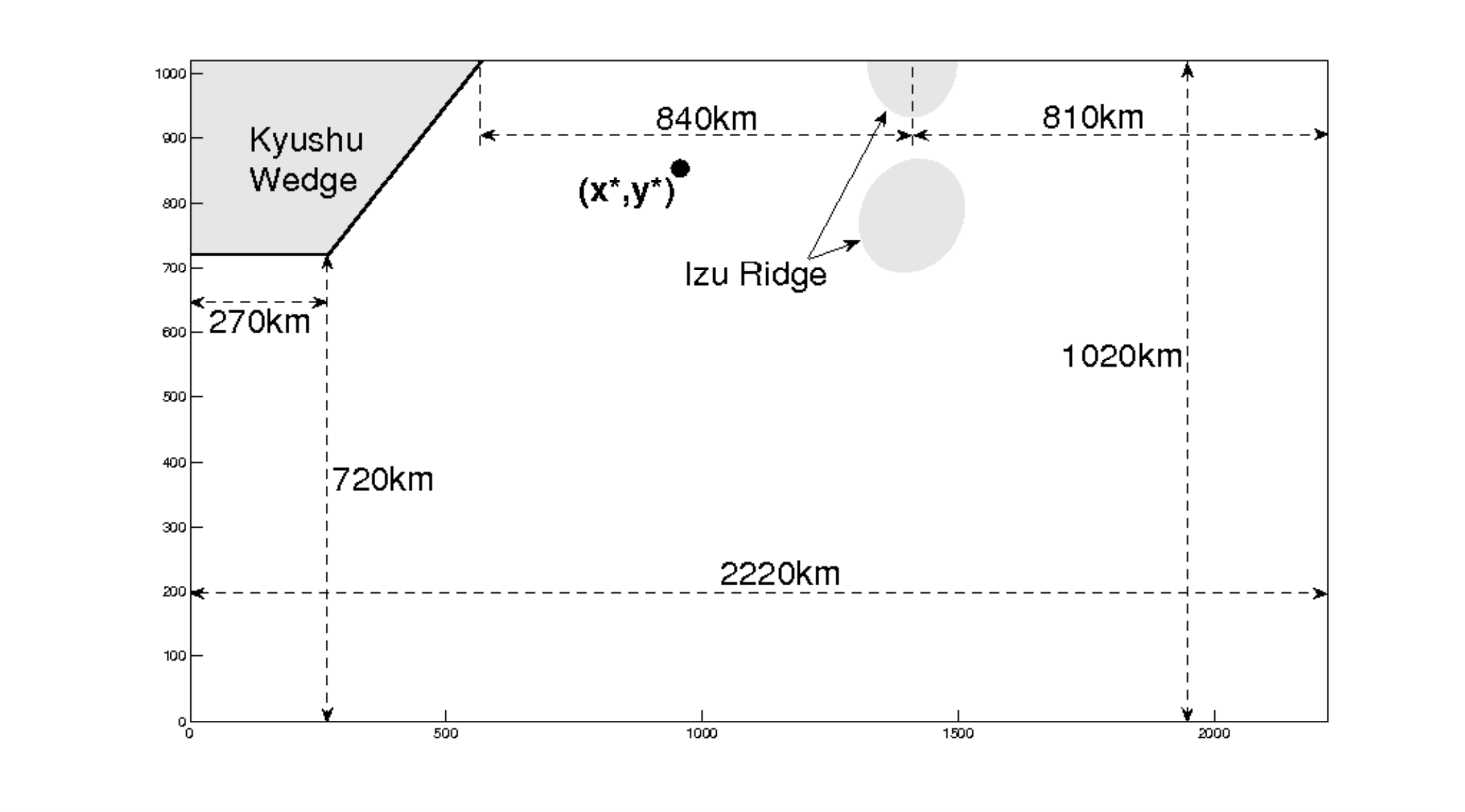}
\caption{   Model geometry.}
\label{domain}
\end{figure}

In the Kuroshio model employed here, as in Chao's original model, the coordinates $ x,y$ are
rotated $20\deg$ counter-clockwise from North-South, and \eqref{spde}
is solved in the domain $D$ shown in Fig.~\ref{domain}.
The viscosity and noise parameters are  set to 
\[
\nu = 8\times 10^{-4} \qquad\text{and}\qquad \sigma = 6 \times
10^{-13}\]
The Coriolis parameter is given by

\[
f = f_0 + f_x x + f_y y
\]
where
$$
f_x = \beta \sin \left(20^\circ\right)\qquad \text{and}\qquad f_y=
\beta\cos\left(20^\circ\right)
$$
and $\beta$ and $f_0$ are given by $\beta = 2\times 10^{-7}$ and $f_0
= 7\times 10^{-5}.$ Unless otherwise specified, all distances are
measured in kilometers and time is measured in seconds.  The function
$r\left(x,y\right)$ is the water depth and away from the two bumps
that model the Izu Ridge, it is set to the fixed value of 1 kilometer.
The northern bump is defined by
\begin{equation*}
r_N\left(x,y\right) = 0.5\
\cos\left(\frac{\pi}{2}\frac{\sqrt{(x-1410)^2+(y-1020)^2}}{90}\right)
\end{equation*}
for
\[
\sqrt{(x-1410)^2+(y-1020)^2}\leq 90
\]
and the southern bump is defined by
$$
r_S\left(x,y\right) = 0.5
\cos\left(\frac{\pi}{4}\sqrt{\left(\frac{x'}{120}\right)^2
+ \left(\frac{y'}{90}\right)^2 }\right)
$$
for
$$
\sqrt{\left(\frac{x'}{120}\right)^2
+ \left(\frac{y'}{90}\right)^2 }         \leq 1
$$
where
\[
x' = \frac{(x-1410) + (y-780)}{\sqrt{2}}
\]
and
\[
y' = \frac{(x-1410) - (y-780)}{\sqrt{2}}.
\]

% \vspace{24pt}

The horizontal velocities $u$ and $v$ satisfy

$$
\left(ru,rv\right)=\left(-\psi_y,\psi_x\right)
$$
where the volume transport streamfunction $\psi$ solves

$$
\mathbf{x} = \frac{\partial}{\partial x} \left(\frac{1}{r}\psi_x\right)
+\frac{\partial}{\partial y}\left(\frac{1}{r}\psi_y\right)
$$

The boundary conditions are
\begin{align*}
&I  & \psi &=0, &\mathbf{x}&=0, &&\text{at}\ y=0,\\
&II & \psi &= -33\,Sv, &\psi_n&=0, &&\text{along the northern boundary},\\
&III  & \psi_x &=0,  &\psi_{xx} &= 0, &&\text{at}\ x=0,\\
&IX  & \psi &= K\left(y\right), &\psi_{xx}&=0, &&\text{at}\ x = 2220
\end{align*}
where 
$$
K(y) = 0\qquad \text{for}\qquad y\leq 870
$$
and
$$
K(y) = -33\,\frac{y-870}{150}\,Sv \qquad\text{for}\qquad y>870.
$$

The temporal and spatial discretization of \eqref{spde} is carried out as follows.
Let $\triangle_x = 30km$ denote the spatial mesh size which is the same 
in both the $x$ and $y$ directions.
For any function $g$ on $D$ define

$$g_{k+\alpha_x,l+\alpha_y} = g\left((k+\alpha_x)\triangle_x,(l+\alpha_y)
\triangle_x\right)$$
for $\alpha_x,\alpha_y\in [-1,1].$
Define the operators
\begin{align*}
&&\delta_x g &= \frac{g_{k+1/2,m}-g_{k-1/2,m}}{\triangle_x},
&\delta_y g &= \frac{g_{k,m+1/2}-g_{k,m-1/2}}{\triangle_x},\\
&&\mu_x g &= \frac{g_{k+1/2,m}+g_{k-1/2,m}}{2},
&\mu_y g &= \frac{g_{k,m+1/2}+g_{k,m-1/2}}{2},\\
&&D_x^0 &= \mu_x\delta_x,
&D_y^0& = \mu_y\delta_y,
\end{align*}
and
\begin{equation}
 L^0 = \delta_x\left( \frac{\delta_x}{r}\right) +
\delta_y\left(\frac{\delta_y}{r}\right).
\end{equation}

  First the equation is discretized  in space using a simple centered difference
scheme which
involves only values of  $\mathbf{x}$ at points 
$\left(k\triangle_x,m\triangle_x\right).$  After replacing $\mathbf{x}$ by its
restriction 
to these points  the system becomes a set of  ordinary stochastic differential
equations,
\begin{equation}\label{sde}
d{\mathbf{x}}_{k,m}(t) = F_{k,m}\left({\mathbf{x}}(t)\right) +\frac{1}{\triangle_x}\sigma\,
d{B}_{k,m}(t)
\end{equation}
where
\begin{multline}\label{drift}
F_{k,m}\left({\mathbf{x}}(t)\right) =   -D^0_x\left(u_{k,j}{\mathbf{x}}_{k,j}\right)
-D_y^0\left(v_{k,m}{\mathbf{x}}_{k,m}\right) \\- 
f_{k,m}\left(\frac{f_x}{f_{k,m}} + D_x^0\left(\frac{1}{r}\right)\right)u_{k,m}
- f_{k,m}\left(\frac{f_y}{f_{k,m}} +
D_y^0\left(\frac{1}{r}\right)\right)v_{k,m}\\
+\nu \left(\delta_x\delta_x + \delta_y\delta_y\right){\mathbf{x}}_{k,m} 
\end{multline}
where
$\psi$ solves
\begin{equation}\label{psi}
L^0 \psi = {\mathbf{x}}
\end{equation}
and

$$
u_{k,m} = -\frac{D_y^0\psi_{k,m}}{r},\qquad v_{k,m} = \frac{D_x^0\psi_{k,m}}{r}.
$$
The $B_{k,m}$ are independent Brownian motions.
Consistent with the previous sections, ${\mathbf{x}}(t_n)$ denotes the solution of
\eqref{discrete}
at the time of $l$th observation.

These stochastic ordinary differential equations cannot be solved explicitly and
therefore require numerical solution.
Here the  spatially discretized system \eqref{sde} is discretized in  time as
\begin{multline}\label{discrete}
\mathbf{x}_{k,m}((n+1)\triangle) = \mathbf{x}_{k,m}(n\triangle) \\+ \left(F_{k,m}(\check{ \mathbf{x}}_{k,m})
+ F_{k,m}(\mathbf{x}_{k,m}(n\triangle)) \right) \frac{\triangle}{2}
+ \frac{\sqrt{\triangle}}{\triangle_x}\sigma\, {\eta}_{k,m}(n)
\end{multline}
where
\begin{equation*}
\check{\mathbf{x}}_{k,m}=\mathbf{x}_{k,m}(n\triangle)+
F_{k,m}\left(\mathbf{x}_{k,m}(n\triangle)\right)\triangle
+ \frac{\sqrt{\triangle}}{\triangle_x}\sigma\, \eta_{k,m}(n)
\end{equation*}
and now for each $k,\ m,$ and $n,$  ${\eta}_{k,m}(n)$ is an independent Gaussian
random variable with mean 0 and variance 1.
The resulting method is adequate for the relatively low Reynolds number 
flow considered here.  A Crank Nicholson type scheme was not applied to the
linear part
of the equation because
the stiffness of the system is not dominated by 
the diffusion term at the level of discretization used here.

\section{Solving for $\widehat \varphi$}
Both Algorithms \ref{pf3} and \ref{ek2} require solving an optimization problem
at least once at each observation time.  As both problems are similar we will
focus on the minimization of functionals of the form
\[
 \int_0^T \frac{1}{2} \left(\dt \varphi(s)-
F(\varphi(s))\right)^\text{\tiny T}(\mathbf{Q}(\varphi(s)))^{-1}\left(\dt \varphi(s)-
F(\varphi(s))\right) ds + g(\varphi(T)).
\]
Note that the minimizer of this action solves the ordinary differential equation
\[
\dt{\widehat \varphi} = F + \sigma \widehat u
\]
where $u\in L^2([0,T])$ minimizes
\[
 \int_0^T \frac{1}{2} u(s)^\text{\tiny T} u(s)  ds + g(\varphi_u(T)).
\]
Because it tends to be well scaled, it is this cost function (after time
discretization) that we choose to optimize.  The discretization corresponding to
the discrete version of the process described in the last appendix is simply
\begin{equation}\label{discaction}
\sum_{n=0}^{T/\triangle-1} \frac{1}{2}u(n)^\text{\tiny T} u(n) \triangle +
g(\varphi_u(T))
\end{equation}
where now
\begin{equation*}
\varphi_u((n+1)\triangle) = \varphi_u(n\triangle)+ \frac{\triangle}{2}
\big(F(\check \varphi_u)
+ F(\varphi_u(n\triangle)) \big) + \frac{1}{\triangle_x} \sigma\, u(n)
\end{equation*}
and
\[
\check{\varphi}_u= \varphi_u(n\triangle)+
F\left(\varphi_u(n\triangle)\right)\triangle
+ \frac{1}{\triangle_x}\sigma\, u(n).
\]

In the minimization of \eqref{discaction} we use the conjugate gradient method
(thereby avoiding the formation of any Jacobian matrices) within a trust region
framework (see in particular the algorithms on pages 69 and 171 in
\cite{nocedalwright}). 
A disadvantage of working in the $u$ variables is that the evaluation of
$\varphi_u(T)$ is a serial operation
making the procedure less amenable to parallel implementation.   
Finally note that Algorithm \ref{pf3} requires repeated minimization of a
function of the form \eqref{discaction} during the generation of each sample
trajectory.  This can be accelerated by the use of a continuation strategy,
using the result of the previous optimization to initiate the optimization at the next time step.

\bibliographystyle{plain}
\bibliography{references}

\end{document}